\documentclass[a4paper,fleqn,usenatbib]{mnras}
\usepackage{newtxtext,newtxmath}
\usepackage[T1]{fontenc}
\usepackage{ae,aecompl}
\usepackage{graphicx}
\usepackage{textcomp}
\usepackage{amsmath}
\usepackage{amssymb}
\usepackage{gensymb}
\usepackage{rotating}
\usepackage{epsfig}
\usepackage{amsmath}
\usepackage{longtable}
\usepackage{color}
 \def\apj{ApJ}
 \def\apjl{ApJL}
 \def\mnras{MNRAS}
 \def\pasp{PASP}

 \def\aap{A\&A}
 
 \def\apjs{ApJS}
 
 \def\nat{Nature}

 \def\mjpb{mJy\,beam$^{-1}$}
 \def\gs{\mathrel{\raise0.35ex\hbox{$\scriptstyle >$}\kern-0.6em\lower0.40ex\hbox{{$\scriptstyle \sim$}}}}
 \def\ls{\mathrel{\raise0.35ex\hbox{$\scriptstyle <$}\kern-0.6em\lower0.40ex\hbox{{$\scriptstyle \sim$}}}}

 \def\Msol{\mathrel{\rm M_{\odot}}}

 \def\Wm2{\,\hbox{W}\,\hbox{m}^{-2}}
 \def\gsim{\mathrel{\raise0.35ex\hbox{$\scriptstyle >$}\kern-0.6em\lower0.40ex\hbox{{$\scriptstyle \sim$}}}}
 \def\lsim{\mathrel{\raise0.35ex\hbox{$\scriptstyle <$}\kern-0.6em\lower0.40ex\hbox{{$\scriptstyle \sim$}}}}
 \def\cgs{\mathrel{\rm erg\,s^{-1}\,cm^{-2}}}

\title[AS2COSMOS]{An ALMA survey of the brightest sub-millimetre sources in the SCUBA-2--COSMOS field}

\author[J.\,M.\ Simpson et al.]{J.\,M.\ Simpson,$^{1,2,3}$
Ian Smail,$^{1}$\thanks{email: ian.smail@durham.ac.uk}
U.\ Dudzevi\v{c}i\={u}t\.{e},$^{1}$
Y.\ Matsuda,$^{3,4}$
B.-C.\ Hsieh,$^{2}$ \newauthor
W.-H.\ Wang,$^{2}$
A.\,M.\ Swinbank,$^{1}$
S.\,M.\ Stach,$^{1}$
F.~X. An,$^{5}$
J.\,E.\ Birkin,$^{1}$
Y.\ Ao,$^{6,7}$
\newauthor
A.\,J.\ Bunker,$^{8}$
S.\,C.\ Chapman,$^{9,10,11}$
Chian-Chou Chen,$^{2}$
K.\,E.\,K.\ Coppin,$^{12}$
S.\ Ikarashi,$^{1}$\newauthor
R.\,J.\ Ivison,$^{13}$
I.\ Mitsuhashi,$^{3,14}$
T.\ Saito,$^{3}$  
H.\ Umehata,$^{15,14}$
R.\ Wang,$^{16}$
Y.\ Zhao$^{17}$
\\
$^{1}$Centre for Extragalactic Astronomy, Department of Physics, Durham University, South Road, Durham DH1 3LE, UK\\
$^{2}$Academia Sinica Institute of Astronomy and Astrophysics, No.\ 1, Sec.\ 4, Roosevelt Rd., Taipei 10617, Taiwan\\
$^{3}$National Astronomical Observatory of Japan, 2-21-1 Osawa, Mitaka, Tokyo, 181-8588, Japan\\
$^{4}$The Graduate University for Advanced Studies (SOKENDAI), Osawa, Mitaka, Tokyo, 181-8588, Japan\\
$^{5}$Inter-University Institute for Data Intensive Astronomy, University of the Western Cape, Robert Sobukwe Road, Bellville 7535, Cape Town, South Africa\\
$^{6}$Purple Mountain Observatory, Chinese Academy of Sciences, Nanjing 210033, People's Republic of China\\
$^{7}$School of Astronomy and Space Science, University of Science and Technology of China, Hefei, Anhui, People's Republic of China\\
$^{8}$Department of Physics, University of Oxford, Keble Road, Oxford OX1 3RH, UK\\
$^{9}$Department of Physics and Astronomy, University of British Columbia, 6225 Agricultural Road, Vancouver, BC V6T 1Z1, Canada \\
$^{10}$National Research Council, Herzberg Astronomy and Astrophysics, 5071 West Saanich Road, Victoria, BC V9E 2E7, Canada \\
$^{11}$Department of Physics and Atmospheric Science, Dalhousie University, Halifax, NS B3H 4R2, Canada\\
$^{12}$Centre for Astrophysics Research, School of Physics, Astronomy and Mathematics, University of Hertfordshire, College Lane, Hatfield AL10 9AB, UK\\
$^{13}$European Southern Observatory, Karl Schwarzschild Strasse 2, D-85748 Garching, Germany\\
$^{14}$Institute of Astronomy, School of Science, The University of Tokyo, 2-21-1 Osawa, Mitaka, Tokyo 181-0015, Japan\\
$^{15}$RIKEN Cluster for Pioneering Research, 2-1 Hirosawa, Wako-shi, Saitama 351-0198, Japan\\
$^{16}$Kavli Institute for Astronomy and Astrophysics, Peking University, Beijing 100871, People's Republic of China\\
$^{17}$Yunnan Observatories, Chinese Academy of Sciences, Kunming 650011,  People's Republic of China\\
}
\date{Accepted ---. Received ---; in original form ---}
\pubyear{2020}

\begin{document}
\label{firstpage}
\pagerange{\pageref{firstpage}--\pageref{lastpage}}
\maketitle

\begin{abstract}
We present an ALMA study of the $\sim$\,180 brightest sources in the SCUBA-2 map of the COSMOS field from the S2COSMOS survey, as a pilot study for AS2COSMOS -- a full survey of the $\sim$\,1,000 sources in this field.
In this pilot we have obtained 870-$\mu$m continuum maps of an essentially complete sample of the brightest 182 sub-millimetre sources ($S_{850\mu \rm m}>$\,6.2\,mJy) in COSMOS. Our ALMA maps detect 260 sub-millimetre galaxies (SMGs) spanning a range in flux density of $S_{870\mu \rm m}$\,=\,0.7--19.2\,mJy. We detect more than one SMG counterpart in 34\,$\pm$\,2\ per cent of sub-millimetre sources, increasing to 53\,$\pm$\,8 per cent for SCUBA-2 sources brighter than  $S_{850\mu \rm m}>$\,12\,mJy.  We estimate that approximately one-third of these SMG--SMG pairs are physically associated (with a higher rate for 
the brighter secondary SMGs,  $S_{870\mu \rm m}\gs$\,3\,mJy), and illustrate this with the serendipitous detection of bright [C{\sc ii}] 157.74\,$\mu$m line emission in  two SMGs, AS2COS\,0001.1\,\&\,0001.2 at $z=$\,4.63, associated with the highest significance single-dish source. Using our source catalogue we construct the interferometric 870-$\mu$m number counts at $S_{870\mu \rm m}>$\,6.2\,mJy. We use the extensive archival data of this field to construct the multiwavelength spectral energy distribution of each AS2COSMOS SMG, and subsequently model this emission with {\sc magphys} to estimate their  photometric redshifts. We find a median photometric redshift for the $S_{870\mu \rm m}>$\,6.2\,mJy AS2COSMOS sample of $z=$\,2.87\,$\pm$\,0.08, and clear evidence for an increase in the median redshift with 870-$\mu$m flux density suggesting strong evolution in the bright-end of the 870\,$\mu$m luminosity function.
\end{abstract}



%
%
%
\section{Introduction}

The brightest high-redshift sources ($S_{870\mu\rm m}\gs $\,10\,mJy) detected in 
sub-millimetre surveys with single-dish telescopes have  far-infrared luminosities of L$_{\rm IR}\gs $\,10$^{13}$\,L$_\odot$, which imply star-formation rates (SFRs) of $\gs $\,10$^3$\,M$_\odot$\,yr$^{-1}$ \citep{Barger14,Dudzeviciute20} and  classify these systems as HyLIRGs (Hyper-luminous InfraRed Galaxies, 
\citealt{RowanRobinson00,RowanRobinson10}).   The immense star-formation rates implied for these systems means that their gas supplies should be rapidly exhausted: $\ls$\,100 Myrs for a typical SMG gas mass of $\sim $\,10$^{11}$\,M$_\odot$ \citep[e.g.,][]{Bothwell13,Birkin20}, and even faster if significant amounts of gas are expelled from the systems by outflows.  This is  $\sim$\,5\% of the length of the era where the activity in the sub-millimetre galaxy (SMGs) population peaks: $z\sim $\,1.8--3.4 \cite[e.g.,][]{Chapman05,Simpson14,Dudzeviciute20}, underlining the potentially short-lived nature and high duty-cycle of these  extreme events.

Although short-lived,   HyLIRG SMGs may represent the most significant individual star-forming events in the Universe, potentially forming an $\sim $\,L$^\ast$ worth of stars in a few 10's Myrs \citep[e.g.,][]{Ivison10,Ivison13}.   Indeed, the intensity of this starburst activity would likely out-radiate all other processes (such as emission from AGN) which can 
 confuse the interpretation of systems with less extreme star-formation rates. Moreover, while extreme, the star-formation processes in these SMGs may be similar to those occurring in a less intense manner in the more numerous bulk of the SMG population, and so their study can aid our understanding of the whole population.   

The number density and physical properties of HyLIRG SMGs, which lie on the rapidly-diminishing tail of high-luminosity sources, are frequently the most challenging for galaxy formation models to reproduce \citep[e.g.,][]{Chakrabarti08, Swinbank08, Dave10, McAlpine19}, and  they can thus provide  strong constraints on  these models. 
However, the reality of many of these extremely luminous SMGs detected in wide-area, but low-resolution, single-dish surveys, has  been called into question.  Strong lensing is clearly responsible for the apparent luminosities of the very brightest sub-millimetre sources, $S_{870\mu\rm m}\gg $\,10--100\,mJy \citep[e.g.,][]{Swinbank10,Ikarashi11,Harrington16}. While at
somewhat fainter fluxes another  concern has arisen from high-resolution interferometric studies, first with SMA and subsequently from ALMA, which suggest that a moderate proportion of bright single-dish sources comprise blends of fainter sources \citep[e.g.,][]{Wang07, Wang11, Younger09, Karim13, Simpson15b, Stach18}.  The low resolution of current single-dish sub-millimetre surveys thus appears to frequently blend several SMGs within a beam to produce a single brighter source, changing the shape of the number counts, most critically by producing a false tail of bright sources, which can also be further boosted by gravitationally lensed sources.   This then complicates the use of these single-dish sub-millimetre counts as an observational constraint on galaxy formation and evolution
models \citep{Cowley15}.

We have undertaken an ALMA continuum survey of bright sub-millimetre sources to investigate these issues, with the goals of  determining the intrinsic form of the bright sub-millimetre counts, better quantifying the influence of blending on single-dish sources, and identifying a sample of intrinsically luminous SMGs to study their physical properties (including the role of any nearby companions in triggering their intense activity).  This pilot study is based on the brightest sub-millimetre
sources  selected from the  SCUBA-2 850-$\mu$m survey of the COSMOS field undertaken by the S2COSMOS project  \citep{Simpson19,An19}.  This ALMA--S2COSMOS (AS2COSMOS) pilot survey represents a systematic programme to obtain, or collate, sub-arcsecond-resolution, sub-millimetre follow-up observations of a complete sample of 850\,$\mu$m-luminous, single-dish-selected sources in this well-studied  field. We will discuss the multi-wavelength properties of these sources in \cite{Ikarashi20} and a sample of serendipitously detected line emitters from our ALMA 
data cubes in \cite{Mitsuhashi20}. Our survey is also complemented by the analysis of all ALMA  archival observations of sources within COSMOS which has been undertaken by Liu, Schinnerer and co-workers \citep{Liu19}.  That study includes a larger sample of sources, but has a more heterogeneous selection (and also mix of ALMA data products) than our study.

The structure of the paper is as follows: in \S 2 we discuss our sample selection, the ALMA observations and our data reduction,
including the construction of our source catalogue and a comparison between the ALMA and SCUBA-2 detections. We also  review the available multiwavelength supporting data. In \S 3 we describe the number counts of sub-millimetre sources we derive, estimate the prevalance of multiple SMGs within SCUBA-2 sources in our survey, including a particularly bright example where we have  serendipitous  confirmation that the two components are associated, and discuss the photometric redshift distribution and trends with sub-millimetre flux in our sample.   Finally in \S 4 we give our conclusions.  We adopt a $\Lambda$CDM cosmology with ${H}_{0}$\,=\,70\,km\,s$^{-1}$\,Mpc$^{-1}$, ${{\rm{\Omega }}}_{{\rm{\Lambda }}}$\,=\,0.7, and ${{\rm{\Omega }}}_{{\rm{m}}}$\,=\,0.3 and, unless otherwise stated, error estimates are from a bootstrap analysis. All magnitudes quoted in our work are in the AB photometric system and we assume a \citet{Chabrier03} initial stellar mass function throughout. 

\section{Observations, Reduction and Analysis}

%
%
\begin{figure*}
    \includegraphics[width=2.0\columnwidth]{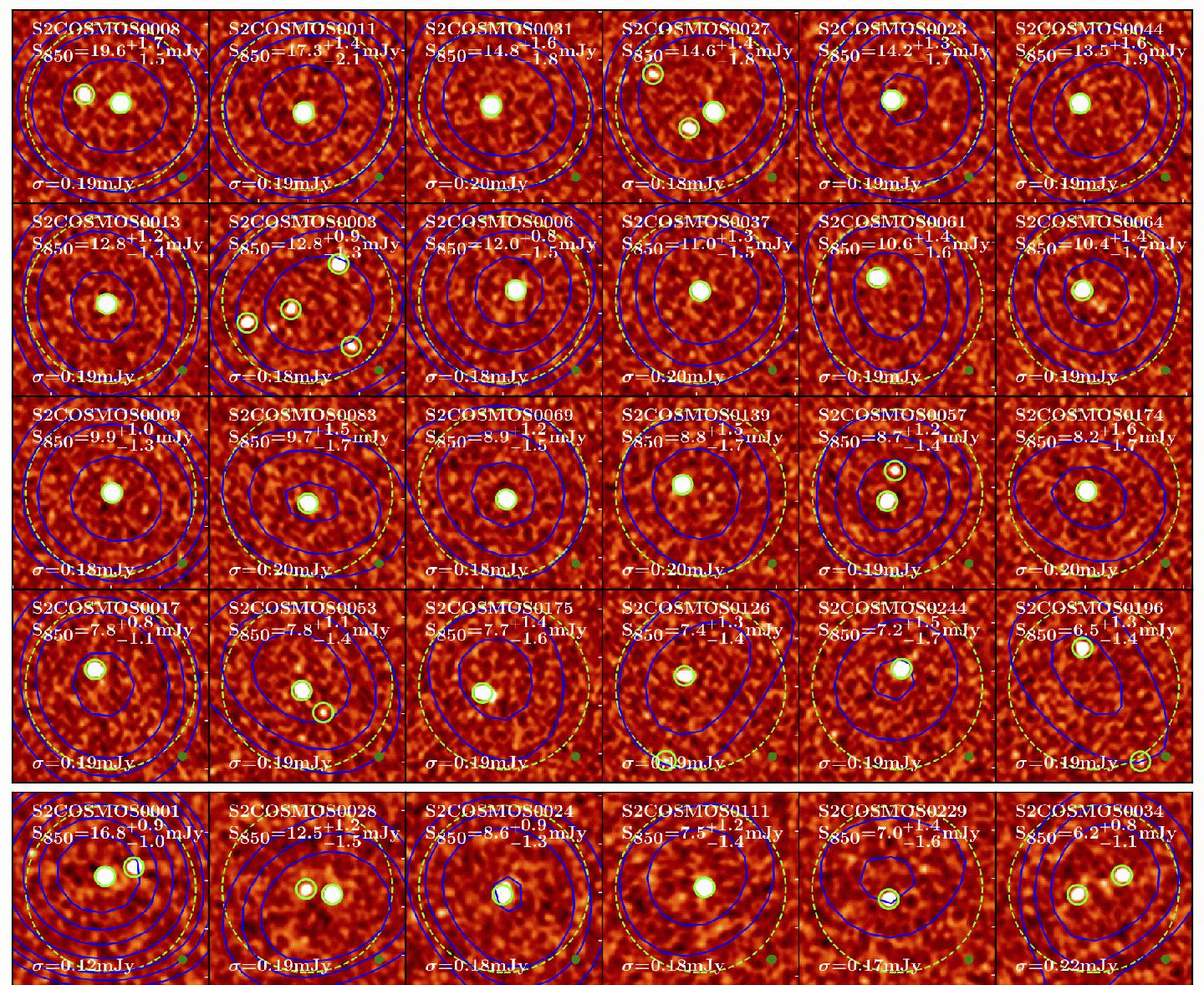}
    \caption{Thirty example ALMA 870-$\mu$m continuum maps from our pilot survey of the 182 brightest SCUBA-2-identified sources in the COSMOS field. The top four rows  were selected at random from our 160 Cycle 4 ALMA targets in bins of single-dish flux density (in descending flux from the top row: $S_{850\mu\rm m}$\,=\,13--20, 10--13, 8--10, 6--8\,mJy). Our AS2COSMOS pilot survey includes 24 archival ALMA maps and we show a randomly-chosen subset of these in the final row to highlight that they have a comparable quality to our Cycle 4 data.  We detect 260 SMGs (circled) at $>$\,4.8\,$\sigma$ across the 182 ALMA 870\,$\mu$m maps, with flux densities of 0.7--19.2\,mJy.   The presence of multiple continuum counterparts in a fraction of the maps is clear (e.g., S2COSMOS\,0003, see \S3.3).
Solid contours represent SCUBA-2 emission at 4, 6, 8, 12, 16, 20, and 24\,$\sigma$. The panels are 20\,$\times$\,20$''$  (160\,$\times$\,160\,kpc at $z\sim $\,2.5) and  a dashed circle represents the 17.3$''$ primary beam of ALMA at 870\,$\mu$m and we show the synthesized beam in the bottom right of each map.
    }
    \label{fig:fieldplan}
\end{figure*}

\subsection{Sample Selection}

The parent sample for our work is selected from a sensitive 850-$\mu$m map of the COSMOS field obtained with SCUBA-2 \citep{Holland13} at the James Clerk Maxwell Telescope (JCMT). This SCUBA-2--COSMOS (S2COSMOS; \citealt{Simpson19}) survey is comprised of two tiers: a {\sc main} region that reaches a median sensitivity of 1.2\,\mjpb over the 1.6\,deg$^{2}$ {\it{Hubble Space Telescope}} ({\it{HST}})\,/\, Advanced Camera for Surveys (ACS) footprint \citep{Koekemoer07}; and a  {\sc supplementary} region that provides an additional 1\,deg$^{2}$ of coverage at a median sensitivity of 1.7\,\mjpb. In this paper we only consider the 1,020 single-dish-identified sources ($S_{850\mu\rm m}$\,=\,2--20\,mJy) that were detected at the $>$\,4\,$\sigma$ significance level in the S2COSMOS {\sc main} survey.

For the ALMA Cycle-4 proposal deadline (April 2016) we employed a preliminary version of the S2COSMOS {\sc main} source catalogue to identify 160 targets for a pilot study into the properties of the most luminous 850\,$\mu$m sources ($S_{850\mu\rm m}$\,$\gsim$\,8\,mJy) in the COSMOS field. Due to a delay in the completion of our ALMA project (see \S\,2.2) and ongoing improvements to the sensitivity of the S2COSMOS map we subsequently adjusted our initial sample selection while retaining the aim of obtaining a flux limited sample of 850\,$\mu$m-luminous sources. As such, in our ALMA Cycle-4 programme we obtained Band 7 imaging for 160 S2COSMOS sources (Figure~\ref{fig:fieldplan}), of which 158 have deboosted/deblended flux densities $S_{850\mu\rm m}$\,$>$\,6.2\,mJy. We note that two SCUBA-2 sources that were observed in the Cycle-4 project scattered to $S_{850\mu\rm m}$\,$<$\,6.2\,mJy in the final S2COSMOS source catalogue ($S_{850\mu\rm m}$\,=\,5.5\,$\pm$\,1.2 and 6.1\,$\pm$\,1.6\,mJy). 

\begin{figure}
    \includegraphics[width=\columnwidth]{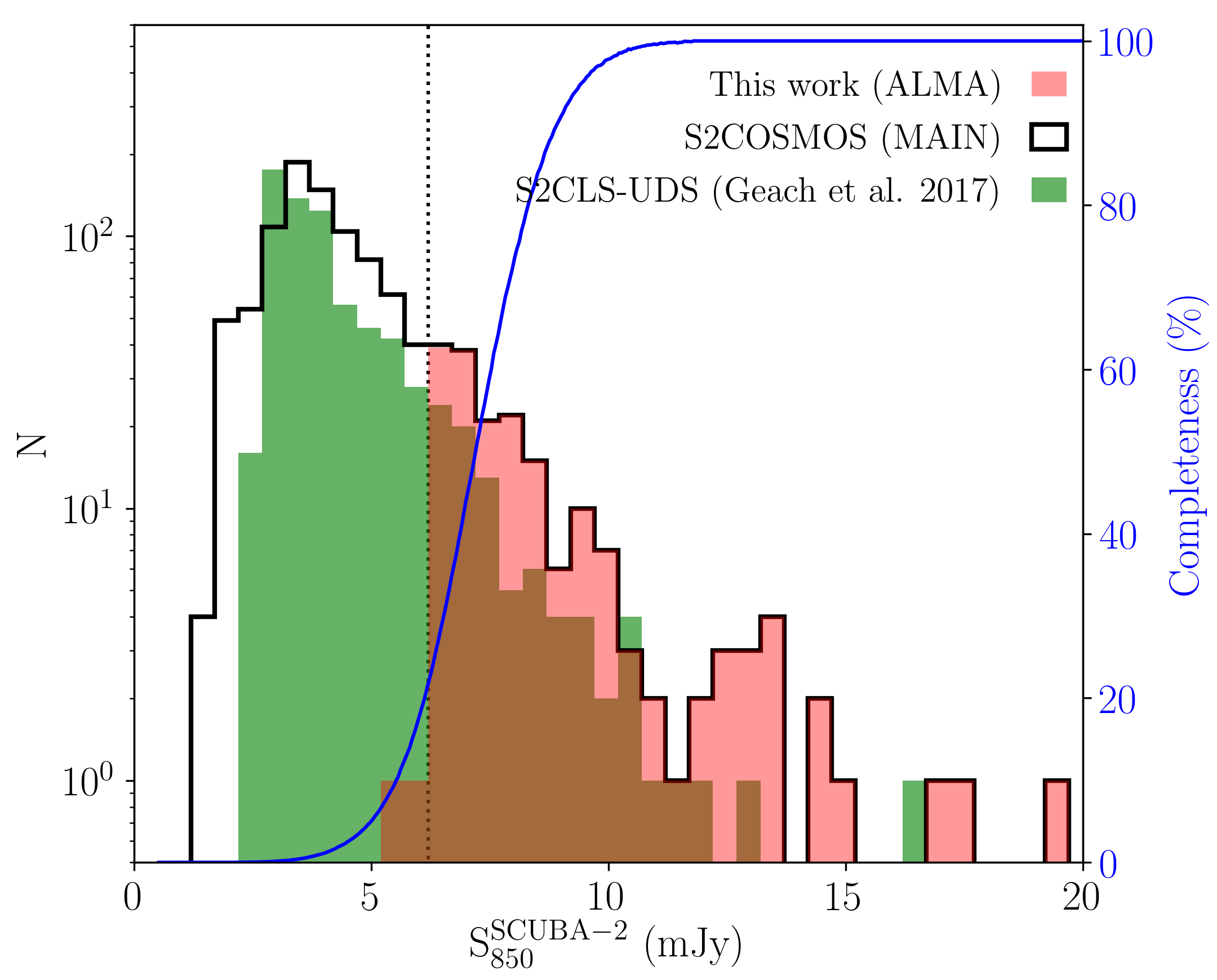}
    \caption{The 850\,$\mu$m flux density distribution of the 182 single-dish-identified sources that comprise our ALMA-S2COSMOS pilot study, compared to the parent distribution of S2COSMOS {\sc main} sources. The AS2COSMOS pilot survey is effectively complete S2COSMOS sources brighter that 6.2\,mJy (182/183; dotted line). For comparison, we show the flux distribution for the S2CLS-UDS sample of SCUBA-2-identified sources \citep{Geach17}, the parent sample for a similar ALMA follow-up study (AS2UDS; \citealt{Stach19}), to demonstrate that AS2COSMOS has roughly twice the source numbers at $S_{850\mu\rm m}$\,$>$6.2\,mJy, relative to a comparable, degree-scale survey. The solid curve represents the effective completeness of the AS2COSMOS pilot survey after accounting for incompleteness in  AS2COSMOS and S2COSMOS studies, as well as the effect of a fixed selection at $S_{850\mu\rm m}$\,$>$\,6.2\,mJy on the S2COSMOS catalogue. AS2COSMOS is estimated to be 22\,per cent, 50\,per cent, and 90\,per cent complete to sources at $S_{850\mu\rm m}$\,=\,6.2, 7.2 and 8.9\,mJy, respectively, across a survey area of 1.6\,deg$^2$. 
    }
    \label{fig:sampleselection}
\end{figure}

The final S2COSMOS {\sc main} source catalogue contains 183 sources with deboosted/deblended 850-$\mu$m flux densities $>$\,6.2\,mJy \citep{Simpson19}. These sources are detected in the S2COSMOS map at a significance ranging from 5.4--28\,$\sigma$ and, as such, we expect the sample to have a false detection rate $\ll$\,1\,per cent \citep{Simpson19}. In our Cycle-4 programme we obtained sensitive 870\,$\mu$m imaging of 158 of these sources and we subsequently identified suitable archival ALMA Band\,7 imaging (see \S\,\ref{subsec:datareduc}) for a further 24. Combining our Cycle-4 observations with the existing archival data means that our AS2COSMOS pilot study is 99.5\,per cent (182\,/\,183) complete for single-dish-identified sources with deboosted/deblended flux densities of $S_{850\mu\rm m}$\,$>$\,6.2\,\mjpb (see Figure\,\ref{fig:sampleselection}), over a survey area of 1.6\,deg$^{2}$. Note that we present our ALMA Cycle-4 maps of two S2COSMOS sources with flux densities $<$\,6.2\,mJy but do not include them in our analysis, where relevant (e.g.\ source counts).   

Finally, we note that there have been a number of prior studies into the properties of far-infrared-luminous sources in the COSMOS field (e.g.\ \citealt{Younger07,Younger09,Smolcic12,Brisbin17,Hill18,Liu19}). A literature search identifies that 45 of the 160 targets in our Cycle-4 ALMA programme have sub-\,/\,millimetre interferometric observations that were typically undertaken with ALMA and\,/\,or PdBI at 1.2--1.3\,millimetre (\citealt{Smolcic12,Brisbin17}), or at $\sim$\,870\,$\mu$m with the SMA (e.g.\ \citealt{Younger09,Hill18}). To ensure that AS2COSMOS represents a homogenous study into the 850\,$\mu$m-luminous population we retained these 45 targets in our ALMA Cycle-4 programme: follow-up observations conducted at a different wavelength to that of the initial sample selection can introduce dust-temperature biases that are challenging to quantify, while the depth of the SMA maps ($\sigma$\,$\sim$\,1--2\,\mjpb) means that the observations are relatively incomplete to sources that lie close to, or below, the flux threshold of our single-dish selection ($S_{850\mu\rm m}$\,$>$\,6.2\,\mjpb). We cross-match our pilot AS2COSMOS sample with these pre-existing catalogues of sub-\,/\,millimetre interferometrically-identified SMGs and provide any alternative identifications in Table~1.

%
%

\subsection{ALMA Data Reduction}\label{subsec:datareduc}
Between 2018 May 15 and 21 we obtained ALMA Band 7 observations of 160 S2COSMOS sources, under project ID: 2016.1.00463.S. Observations were undertaken with a standard correlator set-up for continuum, with four basebands providing 7.5\,GHz bandwidth at a central frequency of 343\,GHz (870\,$\mu$m). For each target, the ALMA pointing centre was fixed to the S2COSMOS source position and, at our observing frequency, the ALMA primary beam (FWHM\,=\,17.3$''$) is well-matched to the SCUBA-2\,/\,JCMT beam (effective FWHM\,=\,14.6$''$; \citealt{Dempsey13}).

Our 160 targets were observed in two ``blocks'' containing 79 and 81 sources, respectively. Each ``block'' was observed twice resulting in a total of four measurement sets. Observations of each ``block'' were conducted with 46 and 48 12\,m antennae, respectively, on baselines ranging from 15--310\,m (median baseline length of 90\,m). Calibrations observations were obtained for each measurement set and same set of calibrators were used throughout. Each measurement set was calibrated in {\sc casa} v\,5.1.1 using the standard reduction pipeline. Phase calibration was conducted using J\,0948+0022, which was observed periodically on a 7\,min cycle, while the absolute flux scale and bandpass were set on J\,1058+0133. We visually inspected the pipeline calibrated data and used the {\sc casa}\,/\,{\sc concat} task to combine the observations of each target into a single measurement set for imaging.

We also include data on a further 24 S2COSMOS sources which were observed in seven publicly-available, archival ALMA projects\,\footnote{Project IDs: 2013.1.00034.S, 2013.1.01292.S, 2015.1.00568.S, 2015.1.01074.S, 2015.1.00137.S, 2016.1.00478.S, 2016.1.01604.S}. To ensure  homogeneity across the AS2COSMOS sample the archival observations were selected on the following criteria, they must have:  a pointing center $<$\,3$''$ from the SCUBA-2 source position (16--84$^{\rm th}$ per centile range from 0.6--2.1$''$); be obtained at an observing frequency of 343\,GHz, and achieve a $1$-$\sigma$ sensitivity of $\lsim$\,0.2\,\mjpb, after applying a taper to broadly match the resolution of our Cycle-4 maps (FWHM\,$\sim$\,0.8$''$). For each of the  archival projects considered here, we retrieved the relevant measurement sets from the ALMA archive and re-ran the data reduction pipeline to fully calibrate the data. Each of the calibrated data sets was visually inspected and any minor issues were corrected (e.g.\ additional channel flagging). 

Imaging the $uv$-data for the AS2COSMOS sources was conducted using {\sc casa} v\,5.1.1, with the new and archival data treated identically. To image each of our science targets we first Fourier transform the $uv$-data to obtain a dirty image, adopting Briggs  weighting (robust parameter\,=\,0.5). Small inhomogeneities in the resolution of the AS2COSMOS maps were accounted for on a map-by-map basis by identifying and applying a two-dimensional Gaussian taper in the $uv$-plane. The appropriate $uv$-taper was chosen such that the synthesized beam of the resulting map has a FWHM\,$\sim$\,0.8$''$ that is well-matched to the resolution of the ``new'' 160 AS2COSMOS observations presented here. We note that for three of the AS2COSMOS targets we cannot construct a $uv$-taper that achieves our target resolution. The $uv$-coverage for the observations of 
S2COSMOS\,0038 and 0111 yields a synthesized beam of 0.95\,$\times$\,0.80$''$, while for S2COSMOS\,0188 we found that further tapering beyond 0.50\,$\times$\,0.47$''$ resulted in a rapid degradation in sensitivity ($\gg$\,0.2\,\mjpb). Overall, the AS2COSMOS pilot sample has a median synthesized beam of 0.80\,$\times$\,0.79$''$ with a variation of $<$\,0.02$''$ across 181 out of 182 maps (see Table\,1).

To clean each of the AS2COSMOS dirty images we use the {\sc tclean} task within {\sc casa} and a two-step procedure. First, the sensitivity of the dirty map was estimated using an iterative, sigma-clipping technique ($\pm$\,4\,$\sigma$). Any sources detected at $\ge$\,6\,$\sigma$ in the dirty image were masked using the {\sc tclean} auto-masking routine, and the masked regions cleaned to 2\,$\sigma$. Note that we enforce that any identified sources are detected at $>$\,4\,$\times$ the expected peak side-lobe level. After the initial clean process has completed we reassess the sensitivity of the map, excluding any masked regions, and perform a second clean. For the second clean process we identify any sources detected at $\ge$\,4.25\,$\sigma$ and clean these to 1\,$\sigma_{870\mu \rm m}$ using the same auto-masking procedure. The resulting maps have a range of 1-$\sigma_{870\mu \rm m}$ depths from 0.11--0.22\,\mjpb (10--90$^{\mathrm{th}}$ per centile $\sigma_{870\mu \rm m}$\,=\,0.18--0.20\,\mjpb) and a median sensitivity of $\sigma_{870\mu \rm m}$\,=\,0.19\,\mjpb (see Table~1). All maps have a pixel scale of 0.1$''$ and a size of 25.6$''$\,$\times$\,25.6$''$.  Representative examples of these data are shown in Figure~\ref{fig:fieldplan}.

%
%
\begin{table*}
 \centering
 \centerline{\sc Table 1: AS2COSMOS Source Catalogue}
\vspace{0.1cm}
 {%
 \begin{tabular}{lcccccccc}
 \hline
 \noalign{\smallskip}
ID & R.A. & Dec. & $S_{\mathrm{SCUBA-2}}$ & Map rms$^{a}$ & Beam & SNR & $S_{\mathrm{ALMA}}^{b}$ & Other ID$^{c}$ \\ 
 & (J2000) & (J2000) & (mJy) & (mJy beam$^{-1}$) & (arcsec) & & (mJy) &  \\ 
\hline \\ [-1.9ex] 
AS2COS0001.1 & 10:00:08.04 & +02:26:12.3 & 16.8$^{+0.9}_{-1.0}$ & 0.12 & 0.80$\times$0.77 & 104.5 & 13.5$^{+0.3}_{-0.3}$ & AzTEC2,COSLA4,AzTECC3a \\ 
AS2COS0001.2 & 10:00:07.84 & +02:26:13.2 & 16.8$^{+0.9}_{-1.0}$ & 0.12 & 0.80$\times$0.77 & 27.8 & 3.6$^{+0.2}_{-0.2}$ & AzTECC3c \\ 
AS2COS0002.1 & 10:00:15.61 & +02:15:49.0 & 13.3$^{+0.7}_{-1.4}$ & 0.12 & 0.80$\times$0.76 & 85.7 & 13.2$^{+0.3}_{-0.2}$ & MM1,COSLA1,AzTECC7 \\ 
AS2COS0003.1 & 10:00:56.95 & +02:20:17.3 & 12.8$^{+0.9}_{-1.3}$ & 0.18 & 0.81$\times$0.79 & 30.5 & 7.5$^{+0.3}_{-0.3}$ & HCOSMOS02.0,131077,AzTECC6a \\ 
AS2COS0003.2 & 10:00:57.57 & +02:20:11.2 & 12.8$^{+0.9}_{-1.3}$ & 0.18 & 0.81$\times$0.79 & 15.0 & 5.1$^{+0.4}_{-0.4}$ & ,HCOSMOS02.1,130891,AzTECC6b \\ 
AS2COS0003.3 & 10:00:57.27 & +02:20:12.6 & 12.8$^{+0.9}_{-1.3}$ & 0.18 & 0.81$\times$0.79 & 10.1 & 2.2$^{+0.3}_{-0.3}$ & HCOSMOS02.4,130933 \\ 
AS2COS0003.4 & 10:00:56.86 & +02:20:08.8 & 12.8$^{+0.9}_{-1.3}$ & 0.18 & 0.81$\times$0.79 & 6.6 & 2.5$^{+0.5}_{-0.5}$ & HCOSMOS02.2,130949 \\ 
AS2COS0004.1 & 10:00:19.75 & +02:32:04.2 & 13.2$^{+0.9}_{-1.1}$ & 0.22 & 0.81$\times$0.79 & 26.3 & 10.8$^{+0.6}_{-0.5}$ & AzTEC5,AzTECC42 \\ 
AS2COS0005.1 & 10:00:23.97 & +02:17:50.1 & 10.3$^{+0.8}_{-1.0}$ & 0.19 & 0.81$\times$0.79 & 29.9 & 8.4$^{+0.4}_{-0.3}$ &  \\ 
AS2COS0005.2 & 10:00:24.03 & +02:17:49.4 & 10.3$^{+0.8}_{-1.0}$ & 0.19 & 0.81$\times$0.79 & 7.5 & 2.1$^{+0.4}_{-0.3}$ &  \\ 
... & ... & ... & ... & ... & ... & ... & ... & ... \\
 \hline\hline \\  [0.5ex]  
 \end{tabular}
 \vspace{-0.5cm}
 \begin{flushleft}
 \footnotesize{The AS2COSMOS source catalogue, showing the sources that are detected in our ALMA maps of the highest significance SCUBA-2 detections across the 1.6\,deg$^{2}$ S2COSMOS {\sc main} survey region. The full catalogue is available in the online journal. $^{a}$ 1--$\sigma$ sensitivity of the non-primary-beam corrected ALMA map $^{b}$ Total flux density, corrected for the ALMA primary beam response $^{c}$ Cross-matched identifications for AS2COSMOS SMGs that have been detected in prior sub-\,/\,mm interferometric observations (see \citealt{Younger07,Younger09,Aravena10,Smolcic12,Bussmann15,Wang16,Brisbin17,Hill18}).}
 \end{flushleft}
}
 \refstepcounter{table}
 \label{table:obs}
 \end{table*}

\subsubsection{Source Extraction}
To construct a source catalogue for our AS2COSMOS pilot survey we first use {\sc sextractor} \citep{Bertin96} to identify any $>$\,2\,$\sigma$ ``peaks'' in the non-primary-beam corrected ALMA maps (Figure~\ref{fig:fieldplan}). At the position of each potential source we measure both the peak flux density and the integrated flux density, using a aperture with a diameter 1.5\,$\times$ the major axis (FWHM) of the synthesized beam. The associated uncertainty on the integrated fluxes is calculated by placing 100 apertures at random on the source-subtracted ALMA maps and taking the standard deviation of the resulting aperture flux densities. 

We expect that our preliminary catalogue of  $>$\,2\,$\sigma$ ``peaks'' is subject to strong contamination from false-detections. To estimate the required significance cut for a robust catalogue of sources we invert the ALMA maps and repeat our source extraction procedure. Within the ALMA primary beam we find that the number of false-detections falls to zero at a peak or aperture significance of $>$\,4.8\,$\sigma$ and $>$\,4.9\,$\sigma$, respectively, and we adopt these criteria here. Applying these selection limits to our preliminary catalogue we obtain a robust sample of 254 SMGs, with each of the 182 ALMA maps containing a minimum of one SMG. 

A visual inspection of the AS2COSMOS maps indicates the presence of potentially-bright sources located marginally outside the ALMA primary beam. Extending our analysis to this region, we find that the false-detection rate falls to zero at a slightly higher peak significance of $>$\,5.1\,$\sigma$, relative to the primary beam area, reflecting the lower data quality in outer parts of each map. We identify six SMGs\footnote{AS2COS\,0015.3, 0055.3, 090.2, 0129.3, 0192.2, and 0196.2} that are located 8.9--11.5$''$ from the phase centre of the relevant map (i.e.\ outside the primary beam) at a peak detection significance of 5.1--10.3\,$\sigma$. These sources are included in our source catalogue and we note that five of the six have a clear counterpart in the available IRAC\,/\,3.6\,$\mu$m imaging of the COSMOS field.

Overall, our pilot survey of 182 S2COSMOS sources yields a sample of 260 AS2COSMOS SMGs with a median detection significance of 24\,$\sigma$ (10--90$^{\mathrm{th}}$ per centile range 6.5--46\,$\sigma$). The brightest SMG in each ALMA pointing is typically located close to the phase centre of the map, with a median offset to the parent SCUBA-2 source of 0.46\,$\pm$\,0.13$''$. Moreover, the median offset in R.A.\ and Dec.\ between the  positions is 0.29\,$\pm$\,0.10$''$ and 0.10\,$\pm$\,0.13$''$, respectively, indicating a good overall level of astrometric agreement between the surveys. In Table\,1 we provide the basic observable properties for each AS2COSMOS source as well as cross-matched identifications for the  69 AS2COSMOS SMGs that have been detected at $\sim$\,arcsec resolution in prior sub-\,/\,millimetre interferometric observations.

\subsection{Flux Estimation}\label{subsec:fluxest}
We estimate the total flux density of each AS2COSMOS SMG by modelling their 870\,$\mu$m emission in the $uv$-plane. High-resolution ALMA imaging of comparable samples of single-dish-identified SMGs indicate that the observed 870\,$\mu$m emission can be well-described by a Sersic profile (e.g.\ \citealt{Hodge16,Gullberg19}), and we adopt that model here. During the fitting procedure we leave five parameters of our model free to vary (R.A., Dec., flux density, half-light radius, and axial ratio) but, given the modest resolution of our ALMA data (FWHM\,$\sim$\,0.8$''$), we fix the Sersic index at $n$\,=\,1, the median best-fit value for 154 of the brighter AS2UDS SMGs observed at 0.15$''$ resolution in ALMA Band\,7 \citep{Gullberg19}. 

Calibrated visibilities for each ALMA target were extracted using {\sc casa} and modelled using a custom written code utilising three publically-available packages. First, we use {\sc proFit} \citep{Robotham17} to construct a pixelated model for all detected SMGs that are detected in a given AS2COSMOS map. This model image is then Fourier transformed into the $uv$-plane using {\sc galario} \citep{Tazzari18}, which yields model visibilities based on the $uv$-coverage of the relevant AS2COSMOS map. Finally, we estimate the best-fit parameters for the input model by minimising the difference between the observed and model visibilities, using $\chi^{2}$ minimisation and the {\sc lmfit} non-linear optimisation suite. False minima in $\chi^{2}$ were mitigated against by repeating the parameter optimisation ten times using randomly-selected starting parameter values, with the iteration at the lowest $\chi^{2}$ value taken as the best-fit solution. 

To estimate the associated uncertainties and characterise any underlying bias on the best-fit flux densities we create 10$^{6}$ simulated ALMA data sets. Each simulated data set is constructed by injecting a single model source into the residual data for a randomly selected AS2COSMOS target. The model source is injected at a random position within a residual map with a Sersic $n=$\,1 light profile that is convolved with the appropriate synthesised beam. The axial ratios and half-light radii of the model sources are drawn at random from a uniform distribution between zero and one, and 0.05$''$ and 0.30$''$, respectively, with the latter chosen to match the  distribution of angular sizes measured for the 154 AS2UDS SMGs from \citet{Gullberg19}. The flux density of each model source is randomly sampled from the parameterised estimate of the sub-\,/\,millimetre counts presented by \cite{Hatsukade18}, with a low flux cut-off at $S_{870\mu \rm m}$\,=\,0.05\,mJy. 

We run our source-detection and visibility-modelling procedures on the simulated data and record the best-fit model parameters for all sources that satisfy our detection criteria. Analysing the results of the simulation we identify the well-known effect of flux boosting, or Eddington bias \citep{Eddington13}, on the recovered flux densities of the simulated sources. Flux boosting describes the statistical overestimation of the flux density of a source detected at a low signal to noise ratio due to the steep shape of the source counts and the effect of random noise fluctuations. On average the recovered flux density of a 4.8\,$\sigma$ source in our simulation is boosted by 15\,per cent, decreasing to a $<$\,4\,per cent bias at $>$\,10\,$\sigma$. To estimate a statistical correction for flux boosting we calculate the median ratio between the recovered and input flux densities of the simulated sources as a function of their detection significance. We use the running median to correct the flux densities of the sources in our AS2COSMOS catalogue, based on their detection significance, and estimate the associated uncertainty on the corrected fluxes from the 1-$\sigma$ scatter in the boosting correction.

Next, we consider the recovered flux density of the simulated sources as a function of input half-light radius. We find that the recovered flux density is unbiased for the average source but do identify a tendency of over\,/\,underestimating the flux densities of sources at smaller\,/\,larger half-light radii. The maximum bias is estimated to be 4\,per cent for sources detected at $<$\,10\,$\sigma$, falling to $<$\,1\,per cent for sources detected at $>$\,15\,$\sigma$ ($S_{870\mu \rm m}$\,$\gsim$\,4.5\,mJy). While we caution that this bias exists for fainter AS2COSMOS source, we stress that our analysis is focused on the bright-end of the SMG population ($S_{870\mu \rm m}$\,$>$\,6.2\,mJy), for which any bias is negligible.

Finally, our parametric estimate for the flux density of each AS2COSMOS SMG may be biased if their 870\,$\mu$m emission is not accurately described by a single Sersic profile. To quantify this effect we  compare the best-fit model flux density for each SMG to the aperture flux density measured during the source detection process. Before making a comparison we must first estimate a correction for the fraction of the total emission that falls outside our adopted aperture. We estimate the aperture correction by creating a stacked radial profile of the emission from all 260 AS2COSMOS SMGs, normalised by their  flux density in an aperture with a diameter 1.5\,$\times$ the beam FWHM (see \S\,\ref{subsec:fluxest}). The resulting profile converges at a radius of $\gsim$\,1.4$''$ with a corresponding  average aperture correction of 1.44\,$\pm$\,0.01. Applying this empirical correction to the aperture flux densities of the AS2COSMOS SMGs we find that agreement with the results of our Sersic profile fitting at the $\pm$\,1.5\,per cent level, on average, confirming that our model fitting procedure is robust on average.

%
%
\begin{figure}
    \includegraphics[width=\columnwidth]{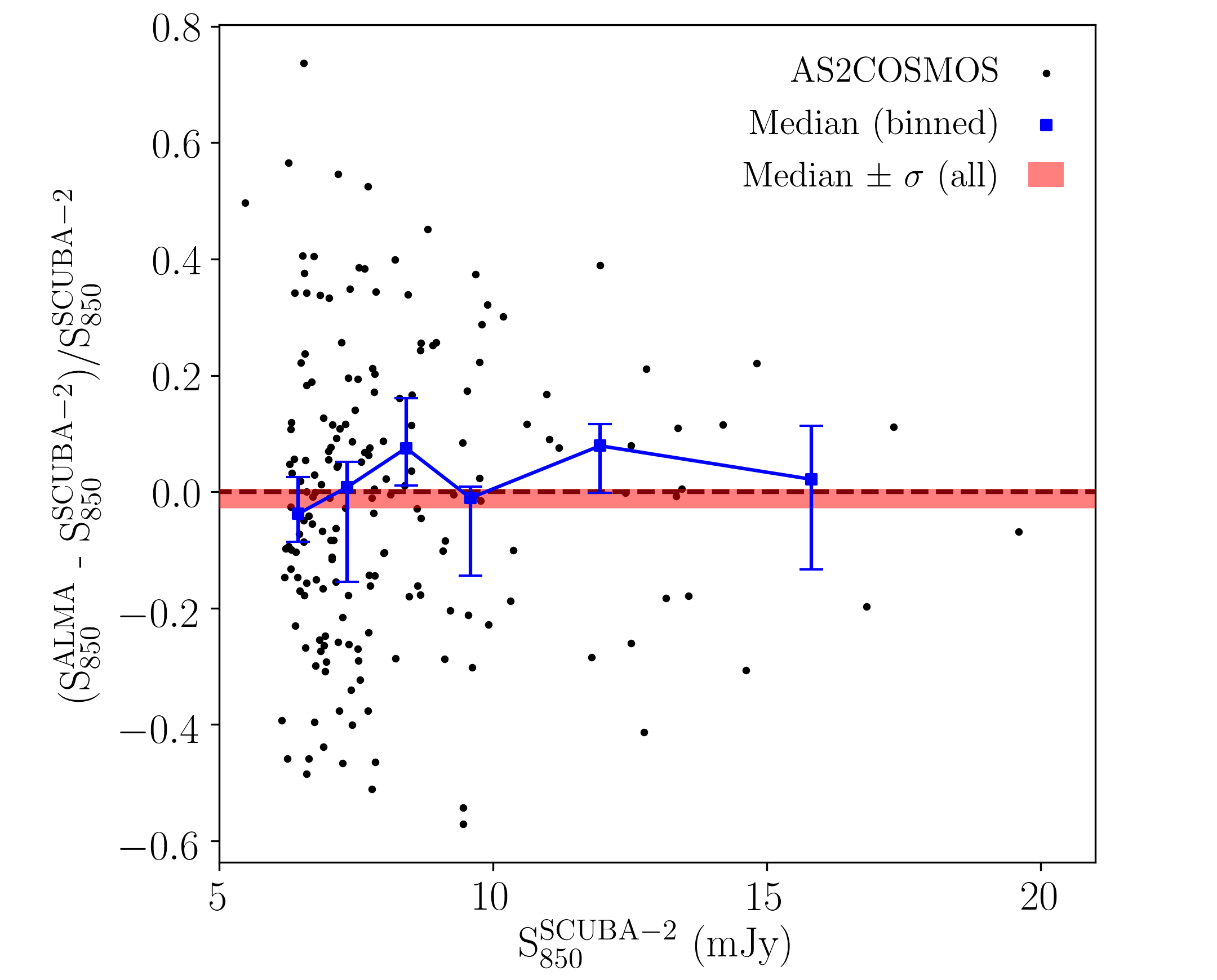}
    \caption{Comparison between the flux density of the brightest component in each AS2COSMOS map and the flux density of the targeted SCUBA-2 source {\it after statistical correction of the latter for boosting and blending}. Overlaid is the median flux recovery and associated uncertainty for all 182 AS2COSMOS maps (shaded), as well as the running median in bins ($\ge$\,0.5\,mJy; solid line) that contain no fewer that 10 sources. We find a statistically-insignificant deficit of $-$\,0.01$^{+0.01}_{-0.02}$\,mJy in the recovered flux density and note that the associated error does not include the expected flux calibration uncertainty of $\sim$\,5\,per cent for both samples.
    }
    \label{fig:flux}
\end{figure}

\subsection{Completeness and Flux Recovery}
\label{subsec:completeness}
Next, we use our Monte Carlo simulations to estimate the completeness level of the AS2COSMOS pilot survey. As expected, the simulations demonstrate that the completeness level of the ALMA maps is strongly dependent on both the angular size and flux density of the input source. If we consider sources with input  flux densities $S_{870\mu \rm m}$\,$>$\,6.2\,mJy, i.e.\ the sample selection for the AS2COSMOS pilot survey, we estimate that our survey is $>$\,99.9\,per cent complete for sources with half-light radii $<$\,1$''$ (this was derived by extending our analysis in the previous section to a broader range in sizes for the SMGs). Prior observations have suggested that typical SMGs have observed  870-$\mu$m half-light radius of $\sim$\,0.1--0.2$''$ \citep{Simpson15b,Ikarashi15,Hodge16,Gullberg19}, as such we consider our catalogue to be complete for SMGs brighter than $S_{870\mu \rm m}$\,=\,6.2\,mJy. 

We stress that our estimate for the completeness level of the AS2COSMOS pilot does not account for any incompleteness in the parent SCUBA-2 sample. The S2COSMOS {\sc main} survey is estimated to be 87\,per cent complete to sources with flux densities of $S_{870\mu \rm m}$\,=\,6.2\,mJy \citep{Simpson19}. Following the procedure detailed in \citet{Simpson19}, we estimate the formal completeness level of the S2COSMOS survey accounting for the sample selection of AS2COSMOS. Accounting for all potential sources of incompleteness we estimate that AS2COSMOS includes 22\,per cent and 90\,per cent of SMGs with intrinsic flux densities of $S_{870\mu \rm m}$\,=\,6.2\,mJy and 9.0\,mJy (see Figure\,\ref{fig:sampleselection}), respectively, that are located within a 1.6\,deg$^{2}$ footprint centred on the COSMOS field. However, we stress that the modest completeness of sources with $S_{870\mu \rm m}$\,=\,6.2\,mJy mostly arises from scattering of sources in a narrow flux range around the adopted flux limit.

Next we consider whether our estimate of the flux recovery is affected by the presence of secondary SMGs in a fraction of the ALMA maps. For each AS2COSMOS map, we create a model image that contains all of the relevant, detected SMGs. We convolve these model maps with an empirical estimate of the SCUBA-2 beam, from the S2COSMOS survey \citep{Simpson19}, and compare the peak flux density of the convolved images to the {\it observed} SCUBA-2 flux densities finding a median ratio of the convolved ALMA-to-SCUBA-2 flux density of 0.94\,$\pm$\,0.01. As expected, the convolved ALMA flux densties are marginally lower than the observed SCUBA-2 fluxes, reflecting that we have not accounted for the effect of flux boosting, or Eddington bias, in the single-dish map. Deboosting corrections for each SCUBA-2 source are provided by \citet{Simpson19}, but these model-dependent corrections account for both Eddington bias and line-of-sight multiplicity. To isolate the Eddington bias correction, we follow the approach in \citet{Simpson19} and 
first construct complete {\it end-to-end} simulations of the S2COSMOS
survey which include both blending of fainter galaxies along the line of sight and the effects of noise
boosting.  We create 100 simulated maps of the S2COSMOS survey following the prescription in \citet{Simpson19}, and with the best-fit parameterisation for the sub-millimetre number counts as the input model (see \S\,\ref{subsec:counts}). Sources were extracted from the simulated S2COSMOS images and we recorded the position and flux of all sub-millimetre emitters ($S_{850\mu\rm m}$\,$\gsim$\,0.05\,mJy) that were injected within 8.7$''$ of each of the recovered SCUBA-2 positions. For each simulated SCUBA-2 source, we identified the corresponding set of sub-millimetre emitters and injected these into a randomly chosen residual map from the AS2COSMOS pilot survey at their model position and flux density. Finally, we ran our source extraction procedure on the simulated ALMA maps cataloging any sources that lie above our threshold for detection and estimating their deboosted flux density. 

These simulations allow us to both test the completeness of our survey, but also assess the contribution
of clustering on the presence of multiple SMG counterparts to a single-dish source in  \S\,\ref{subsec:mult}.  Here we use the SCUBA-2--ALMA simulations to estimate the fraction of each deboosting/deblending correction factor provided by \citep{Simpson19} that arises from line-of-sight multiplicity in the single-dish population. Accounting for this line-of-sight multiplicity correction on the deboosted S2COSMOS flux densties, we find a median ratio of convolved ALMA to deboosted SCUBA-2 flux density of 0.99\,$\pm$\,0.01, indicating a good level of agreement between the flux scales of the two surveys.

We can also compare the fluxes of the {\it brightest} component in each ALMA map to the estimated flux for
that source from \citep{Simpson19} which included statistical corrections for both blending and noise boosting. We note that the statistical correction for line-of-sight multiplicity assumes
no clustering, although it appears that physically-associated SMG-SMG pairs do not dominate in the overall AS2COSMOS sample (see \S\,\ref{subsec:mult}). We find that flux density of the brightest SMG in each ALMA map is, on average, 1$^{+1}_{-2}$\,per cent lower than the flux density of the corrected SCUBA-2 source (see Figure\,\ref{fig:flux}) which shows good agreement.  This suggests that with knowledge of the {\it true form} of the SMG number counts, it is possible to statistically correct for the effects of both multiplicity and boosting in single-dish counts to estimate the true sub-millimetre brightness of the counterparts.

In summary, we conclude that the AS2COSMOS SMGs account for the bulk of the emission traced by each of the targeted SCUBA-2 sources, although we reiterate that there is a nominal flux calibration uncertainty of $\sim$\,5\,per cent on both flux scales.

\subsection{Archival multiwavelength observations}
\begin{figure*}
    \includegraphics[width=2\columnwidth]{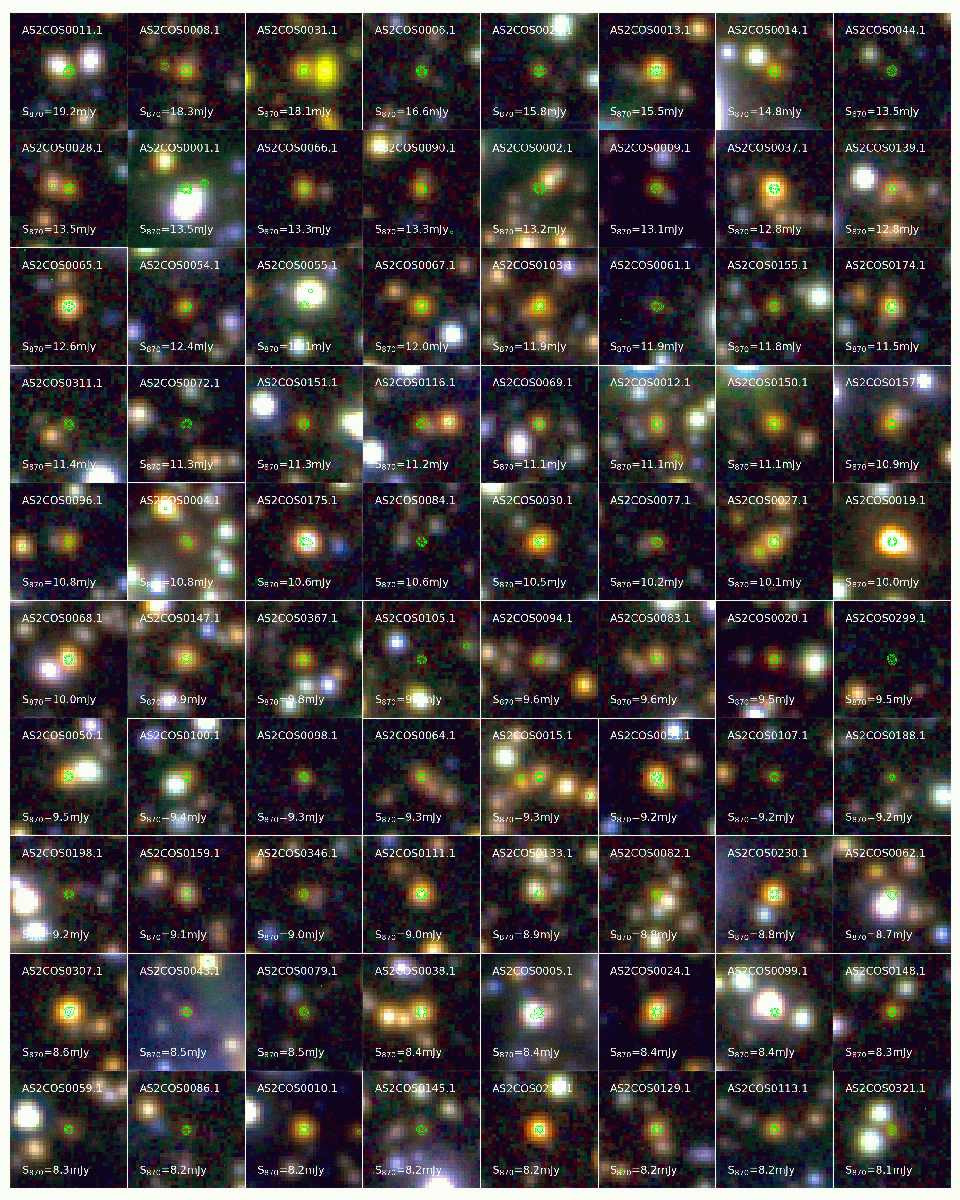}
    \caption{20$''$\,$\times$\,20$''$ true colour images ($K_{s}$, 3.6\,$\mu$m \& 4.5\,$\mu$m) of the 80 brightest 870-$\mu$m SMGs ($S_{870\mu \rm m}$\,$\ge$\,8.1\,mJy) in the AS2COSMOS sample (the remaining SMGs are shown in Fig.~A1). Each image is centred on the ALMA source position and are ordered by decreasing ALMA 870-$\mu$m flux density. Contours represent the ALMA 870\,$\mu$m detections are are overlaid at 4, 10, 20, and 50\,$\sigma$. These images demonstrate that the AS2COSMOS SMGs are typically very red and\,/\,or faint at near-to-mid infrared wavelengths, relative to the field population. We find that 9\,$\pm$\,1\,per cent of the AS2COSMOS SMGs are not detected in the Ultravista\,/\,$K_{s}$ imaging, at the $>$\,5\,$\sigma$ significance level, but note that all but one of these sources are detected in the deblended IRAC imaging (median $m_{4.5}$\,=\,23.5\,$\pm$\,0.2). Overall, the AS2COSMOS SMGs have a median $K_{s}-$\,4.5\,$\mu$m colour of 1.24\,$\pm$\,0.04\,mag, reflecting the importance of sensitive mid-infrared imaging for conducting a unbiased study in to the stellar emission of 870\,$\mu$m-luminous sources.
    }
    \label{fig:multimages}
\end{figure*}

The COSMOS field has been the target of numerous imaging campaigns at X-ray-to-radio wavelengths, and has one of the deepest sets of multi-wavelength data available over a degree-scale area. We make use of this extensive imaging to construct the UV-to-radio SED of each AS2COSMOS SMG, which we subsequently model to derive their physical properties (e.g.\ photometric redshifts, far-infrared luminosities) in \cite{Ikarashi20}. The following describes the data sets used in our analysis and the methods used to determine the multi-wavelength photometry of the AS2COSMOS SMGs.  We use these data in Figures~\ref{fig:multimages} and \ref{fig:multimages_app} to illustrate the appearance of the SMGs in our sample in the observed near-/mid-infrared wavebands.

%
%

\subsubsection{Optical-to-Near-infrared imaging}
The COSMOS2015 catalogue \citep{Laigle16} includes 27-band, optical-to-near-infrared photometry for near-infrared-selected sources in the COSMOS field. \citet{Laigle16} homogenise the $u$-to-$K_{s}$-band imaging (FWHM\,$\sim$\,0.5--1.0$''$) to a broadly consistent PSF and identify sources in a stacked $zYJHK_{S}$ ``detection'' image. For each detected source, flux densities are extracted in a 3$''$ diameter aperture on the PSF-homogenised images and aperture corrected to a total flux measurement.

Before cross-matching the AS2COSMOS and COSMOS2015 catalogues we estimate the probability of a false-match as a function of matching radius. We construct a catalogue of 10$^{4}$ random positions across the field and cross match these with the COSMOS2015 source positions. The probability of a false-match is estimated at $\sim$\,6.6\,per cent at a separation of 0.85$''$ and we adopt this as our matching radius. Cross-matching the AS2COSMOS and COSMOS2015 catalogues we identify 179\,/\,260 matches within 0.85$''$, with a median separation of 0.19$''$ (false-match probability $\sim$\,0.4\,per cent). Note that we correct a small astrometric offset between the catalogues of 0.08\,$\pm$\,0.01$''$ in R.A., but do not find a significant offset in Declination. We comment that the three SMGs that are offset by 0.70--0.85$''$ to an associated optical counterpart appear morphologically-complex and\,/\,or faint in the $K$-band imaging, consistent with the expectation that there will be significant systematic offsets ($\sigma$\,$\sim$\,0.3$''$; \citealt{Chen15}) between the rest-frame far-infrared and  optical emission in these heavily obscured sources (see Figure~\ref{fig:multimages}).

The deep $YJHK_{s}$ imaging provided in the COSMOS2015 catalogue is derived from the second data release (DR2) of the Ultravista survey \citep{McCracken12}. The fourth data release from the Ultravista survey provides imaging (FWHM\,=\,0.8$''$) that is up to $\sim$\,0.5\,mag deeper than the earlier DR2 imaging; the DR4 imaging has a limiting depth of $K_{\mathrm{s}}$\,=\,26.4--25.4\,mag and $K_{\mathrm{s}}$\,25.3--25.1\,mag (3\,$\sigma$ depth in 2$''$ diameter aperture) in four ultra-deep and deep stripes, respectively. To improve upon the near-infrared photometry of the AS2COSMOS sources we replace the $YJHK_{s}$ COSMOS2015 photometry with 2$''$ aperture photometry extracted at the position of each SMG on the DR4 imaging. The associated background level and uncertainty on our aperture flux densities are estimated in a 1$'$\,$\times$\,1$'$ region centred on each SMG. To convert our aperture flux densities to a total flux density we calibrate our results to those in the COSMOS2015 catalogue; for the SMGs with a counterpart in the COSMOS2015 we determine the median ratio between the DR4 aperture flux density and the COSMOS2015 total flux and apply this as an aperture correction to our measurements. Note that we visually inspect the near-infrared imaging and discard photometry for 18 SMGs where the aperture flux are strongly contaminated by a neighbouring, likely to be foreground, source.  This leaves us with a sample of 232 SMGs which have detectable emission above 3-$\sigma$ in the $K_S$-band.

In addition, the second data release \citep{Aihara19} of Hyper-SuprimeCam (HSC) Subaru Strategic Program (SSP) provides $g$, $r$, $i$, $z$, and $Y$ imaging ($\sim$\,0.6$''$ seeing) of the COSMOS field at a 3-$\sigma$ equivalent depth of 28.1, 27.9, 27.9, 27.4, and 26.4\,mag, respectively. This imaging reaches $\sim$\,1\,mag deeper than the optical imaging used in the COSMOS2015 catalogue and we include it in our analysis. The HSC-SSP data release provides aperture corrected flux densities (2$''$ diameter aperture) for all sources detected at $\ge$\,5$\sigma$ in any of the $g$, $r$, $i$, $z$, or $Y$ images. We cross-match the HSC-SSP catalogue to the AS2COSMOS source positions, adopting the same 0.85$''$ matching-radius. This yields 158 optically-detected counterparts to the AS2COMSOS SMGs, 20 of which are not present in the COSMOS2015 catalogue.  Where a source lacks an entry in the HSC-SSP catalogue we visually assess the cause using $gri$ thumbnail images from the HSC archive and determine if the source is undetected, in which case we adopt the appropriate magnitude limits, or if it is blended with or obscured by a bright nearby source (where we remove
the photometry -- although this only applies to $\sim$\,10 sources and we confirm 
that it doesn't influence their best-fit SEDs derived in \S 2.6).

\subsubsection{Mid-infrared imaging}
Mid-infrared imaging at 3.6, 4.5, 5.8, 8.0\,$\mu$m of the COSMOS field is provided by the {\it Spitzer} Large Area Survey with Hyper-SuprimeCam (SPLASH; see \citealt{Steinhardt14,Laigle16}). The SPLASH imaging is comprised of data that was obtained with the Infrared Array Camera (IRAC; \citealt{Fazio04}) on board the {\it Spitzer Space Telescope} as part of the SPLASH-COSMOS, {\it Spitzer}-COSMOS (S-COSMOS), {\it Spitzer} Extended Deep Survey (SEDS), and {\it Spitzer}-CANDELS datasets, and provides coverage at 3.6--8.0\,$\mu$m for all AS2COSMOS SMGs. 
The IRAC data reach an average 3\,$\sigma$ limiting magnitude for point sources of 23.9, 23.6, 22.5, and 22.0\,mag at 3.6, 4.5, 5.8, 8.0\,$\mu$m, respectively. 

The resolution of the IRAC imaging (FWHM\,$\sim$\,2$''$) is significantly coarser than the optical-to-near-infrared imaging of the field, and more sophisticated methods than simple aperture photometry are required to derive accurate flux densities for the AS2COSMOS SMGs. For the COSMOS2015 catalogue, deblended IRAC flux densities were determined for all optically-selected sources using {\sc iraclean} \citep{Hseih12}. Briefly, {\sc iraclean} deblends the IRAC imaging using a higher resolution image as a prior, in this case, the stacked $zYJHK_{S}$ ``detection'' image. The IRAC PSF is constructed dynamically across the field and each image is deblended following a process that is functionally identical to {\sc clean} deconvolution in radio interferometry. To estimate deblended IRAC photometry for the AS2COSMOS SMGs we again use {\sc iraclean} but update the prior to include all AS2COSMOS SMGs, including those not formally detected in the $zYJHK_{S}$ stack. We follow an identical deblending procedure to that described in \citet{Laigle16}, and refer the reader to that work for further details (see also \citealt{Hseih12}).  The red colours of the SMG population means that these bands provide the highest detection rate for our targets, with 238 of the 260 sources  detected at 4.5-$\mu$m.

\subsubsection{Far-infrared imaging}
Far-infrared imaging of the COSMOS field was obtained at 24\,$\mu$m with the Multiband Imaging Photometer (MIPS) on board the {\it Spitzer Space Telescope}, and at 100\,$\mu$m and 160\,$\mu$m with the Photodetector Array Camera and Spectrometer (PACS; \citealt{Poglitsch10}) on board the {\it Herschel Space Observatory}. The 24\,$\mu$m data is taken from the COSMOS-{\it{Spitzer}} programme \citet{Sanders07} and reaches a 1-$\sigma$ depth of $\sim$\,15\,$\mu$Jy \citep{LeFloch09}. The 100\,$\mu$m and 160\,$\mu$m imaging was obtained as part of the PACS Evolutionary Probe (PEP) survey has a 1-$\sigma$ sensitivity of $\sim$\,1.4\,$\mu$Jy and $\sim$\,3.5\,$\mu$Jy, respectively (see \citealt{Berta11}, but also \citealt{Jin18}).

Source confusion is a concern in the low-resolution MIPS and PACS imaging (FWHM\,=\,6--12$''$) and to estimate accurate flux densities for individual sources the emission in the maps must be deblended based on prior catalogue lists. In this work we primarily utilise the ``super-deblended'' catalogue presented by \citet{Jin18}, which contains deblended 24--160\,$\mu$m photometry for $K_{s}$- and 3\,GHz-selected sources in the COSMOS field. Briefly, \citet{Jin18} deblend the MIPS and PACS imaging of the field by PSF-fitting at the position of 194,428 sources in their prior catalogue following the methodology presented by \citet{Liu18}. Cross-matching the optical and radio counterparts to the AS2COSMOS SMGs with the \citet{Jin18} catalogue yields 24--160\,$\mu$m photometry for 228\,/\,260 AS2COSMOS SMGs. 

The source catalogue presented by \citet{Jin18} is incomplete to far-infrared-luminous, but radio- and\,/\,or $K_{s}$-faint, sources. To increase the completeness level of our 24--100\,$\mu$m photometry we also match to the source catalogue from the PACS\,/\,PEP survey (\citealt{Lutz11}; see also \citealt{Magnelli13}) that was constructed using a 24\,$\mu$m-only prior list. We find an additional seven counterparts to the AS2COSMOS SMGs, within a matching radius of 2$''$ (see \citealt{Chen16a}), increasing the overall completeness level for the AS2COSMOS SMGs to 235\,/\,260.

Imaging at 250, 350 and 500\,$\mu$m of COSMOS was taken with the Spectral and Photometric Imaging Receiver (SPIRE; \citealt{Griffin10}) on board the {\it Herschel Space Observatory} as part of the {\it Herschel} Multi-tiered Extragalactic Survey (HerMES; \citealt{Oliver12}). These data are particularly important for our analysis as they are expected to sample the peak of the rest-frame dust emission from the AS2COSMOS SMGs ($\lambda_{\mathrm{obs}}$\,$\sim$\,300\,$\mu$m for a source with a characteristic dust temperature of 30\,K at $z$\,$\sim$\,2.5) and, as such, constrain the total infrared luminosities of our sample. Due to the coarse resolution of the {\it Herschel} SPIRE imaging (FWHM\,=\,18--35$''$) source deblending is again crucial for determining accurate flux densities for each of the AS2COSMOS SMGs.

We deblend the 250, 350 and 500\,$\mu$m imaging following the method described in \citet{Swinbank14}. Briefly, we construct a prior list of MIPS\,/\,24\,$\mu$m, VLA\,/\,3\,GHz (see \S\,2.6), and ALMA\,/\,870\,$\mu$m sources that are used to deblend the low-resolution maps. The typical $K_{s}$-selected sources included by \citet{Jin18} in the deblending of the 24--160\,$\mu$m imaging are not expected to be luminous in the SPIRE imaging and, as such, we omit these from our prior list. Deblending of the SPIRE maps is achieved by PSF-fitting to the observed flux densities at the position of all sources in the prior catalogue. To avoid ``over-deblending'', the maps were deblended in order of increasing wavelength with only ALMA SMGs and\,/\,or sources detected at $>$\,2\,$\sigma$ in the proceeding map retained in the prior list. The associated uncertainties on the deblended flux densities, and detection limits of the SPIRE maps, were determined through extensive Monte Carlo simulations to inject and recover simulated sources in each map. We find that the deblended 250, 350 and 500\,$\mu$m maps reach a typical $3$\,$\sigma$ limit for detection of 7.4, 8.1, and 10.6\,mJy, respectively. 

Overall, we find that 235\,/\,260 of the AS2COSMOS SMGs are detected in at least one wave-band between 24--500\,$\mu$m, with 222\,/\,260 (85\,per cent) in at least one band between 100--500\,$\mu$m. To first order, the high detection fraction for the AS2COSMOS SMGs at 100--500\,$\mu$m reflects our selection of bright single-dish sources for ALMA follow-up observations. Indeed, the 38 SMGs without a detection in either the PACS or SPIRE imaging have a median 870\,$\mu$m flux density of $S_{870\mu \rm m}$\,=\,4.1\,$\pm$\,0.5\,mJy, significantly lower than the median flux density of $S_{870\mu \rm m}$\,=\,7.1\,$\pm$\,0.2\,mJy for the ``detected'' subset.

\subsubsection{Radio}
To analyse the radio properties of the AS2COSMOS SMGs we utilise deep 3\,GHz imaging of COSMOS undertaken in a Large Project with the Karl. G. Jansky Very Large Array \citep{Smolcic17}. Briefly, the 3\,GHz map of the COSMOS field reaches a median sensitivity of 2.3\,$\mu$Jy, at a resolution of 0.75$''$, over 2\,deg$^{2}$. In the following, we use the source catalogue presented by \cite{Smolcic17}, which contains total flux densities for 10,830 sources that were identified at the $>$\,5\,$\sigma$ significance level.

We cross-correlate the AS2COSMOS and VLA\,/\,3\,GHz catalogues and identify 191 counterparts to the AS2COSMOS SMGs within a matching radius of 1$''$. The adopted matching radius is comparable to that used to identify optical counterparts to each of the SMGs and, considering random positions in the field, we estimate a false-match probability of $\sim$\,0.1\,per cent. Note that extending the matching radius to 2$''$ does not yield any further unique 3\,GHz counterparts. A visual inspection the VLA imaging shows that two pairs of SMGs with a small on-sky separation ($\sim$\,1--2$''$) have distinct, well-separated, peaks (SNR\,=\,18--33) in the 3\,GHz map, but are grouped into a single source in the VLA\,/\,3\,GHz catalogue. To obtain deblended 3\,GHz flux densities for these SMGs (AS2COS\,0051.1\,/\,.2 and 0228.1\,/\,.2) we use the {\sc casa}\,/\,{\sc imfit} routine to simultaneously model the emission from
each pair of components. Overall, we identify 3\,GHz counterparts to 193\,/\,260 AS2COSMOS SMG with flux densities of 12--650\,$\mu$Jy. 

To investigate the radio properties of the 3\,GHz-faint SMGs we stack thumbnails extracted from the VLA map at their positions. These 67 SMGs are detected at the 27\,$\sigma$ level in the stacked image with an average peak flux density of 8.1\,$\pm$\,0.3\,$\mu$Jy, placing the average source marginally below the formal limit for detection in the VLA map ($\sim$\,11.5\,$\mu$Jy). Motivated by the strength of the stacked emission, we estimate the 3\,GHz flux density of each of the radio-faint SMGs by extracting the pixel flux density at the position of each source in the VLA map: the 3\,GHz maps are calibrated in units of $\mu$Jy per beam and the pixel value represents the total flux density for an unresolved source at a given position. At the resolution of the VLA imaging we expect that the radio emission from the AS2COSMOS SMGs will be marginally-resolved (intrinsic FWHM\,$\sim$\,0.6$''$; e.g.\ \citealt{Biggs08,Miettinen15,Thomson19}) and, as such, our simple flux estimates will systematically underestimate the total flux of each source. To correct for this effect we compare the pixel and total flux densities the 193 AS2COSMOS with counterparts in the VLA\,/\,3\,GHz catalogue. We determine a median total-to-peak flux density ratio of 1.21\,$\pm$\,0.03 for the average SMG, which we use to correct our flux estimates for the 67 3\,GHz-faint SMGs to a total flux density.

\subsubsection{X-ray}
The {\it Chandra} COSMOS Legacy Survey \citep{Civano16} provides coverage of the AS2COSMOS SMGs in the 0.5--2\,keV (soft) and 2--10\,keV (hard) bands at an effective exposure of 160\,ksec across our full survey area. The source catalogue for the survey contains 4,016 point sources that are detected in any combination of the soft, hard and full (0.5--10\,keV) bands (flux limit of 8.9\,$\times$\,10$^{-16}$\,$\cgs$ in the full band). 

Matching the {\it Chandra} and ALMA source catalogues within the 3\,$\sigma$ positional uncertainties on the X-ray positions we identify 24 counterparts to the AS2COSMOS SMGs, at a median positional offset of 0.56\,$\pm$\,0.08$''$.  Note that we choose to include a ``match'' to AS2COSMOS308.1 despite the X-ray source lying offset to the ALMA position at the 4.8\,$\sigma$ significance level. The offset between the X-ray and ALMA positions is 0.84$''$ (i.e. 1\,/\,3 of the {\it Chandra} beam) and a visual inspection of the optical-to-near-infrared imaging indicates that there is no clear alternative counterpart to the source at optical-to-radio wavelengths.  

We also cross-correlated the AS2COSMOS sources with the {\it XMM-Newton} X-ray survey of COSMOS by \cite{Brusa10}, but find no additional reliable identifications.

\subsection{Panchromatic SED fitting}

We first summarise the detection rates for the AS2COSMOS SMGs in the various wavebands discussed above.  Other than at 870-$\mu$m, the highest detection rate is found in the 
 {\it Spitzer} IRAC bands with 238 of the 260 SMGs with $\geq\,$3-$\sigma$ detections in the  4.5-$\mu$m band.  The detection rate drops markedly in bluer wavebands, as has been seen for
previous studies of this dusty and typical high-redshift population, with 196 sources and 103 being detected above 3-$\sigma$  in the $Y$ and  $B$-bands respectively.   While at longer wavelengths, 174 of the 260 SMGs have $\geq$\,3-$\sigma$
detections at 350-$\mu$m  from the deblended photometry.  Overall the median number of photometric constraints on the SMG's SEDs is 18 bands, with the maximum being 24 and just five SMGs have five-or-fewer bands used to constrain
their SEDs.   The detection rates in $K_S$, 3.6\,$\mu$m and 350\,$\mu$m of 89, 92 and 67 per cent are slightly higher than the
corresponding values 83, 90 and 59 per cent for the AS2UDS study of \cite{Dudzeviciute20} to which we compare our results.

Having collated the multiwavelength observations of our ALMA-identified SMG sample we now use the {\sc magphys} spectral energy distribution (SED) modelling programme \citep{daCunha08,daCunha15,Battisti19} to fit the multiwavelength SEDs of these sources.  Our approach here is to match  the method used in the analysis of $\sim$\,700 ALMA-identified SMGs from the AS2UDS survey by \cite{Dudzeviciute20}, to allow us to easily compare the physical properties estimated by {\sc magphys} for that sample to the typically more luminous sources studied here.

We refer the reader to \cite{Dudzeviciute20} for a detailed description of the application and testing of the {\sc magphys} software on large samples of observed and theoretical galaxy SEDs, with a particular focus on dust-obscured star-forming galaxies.  These include testing of the precision of the derived photometric redshifts using a sample of 
$\sim$\,7,000 spectroscopically-identified galaxies at $z\sim$\,0--5 in the UKIDSS UDS field, including 44 SMGs, and assessment of the systematic uncertainties in other physical parameters from the model through its application to model galaxy SEDs for strongly star-forming galaxies derived from the {\sc eagle} simulation \citep{McAlpine19}. From the spectroscopic comparison they determine  $\Delta z$/\,$(1+z_{\rm
spec})\,$\,$= -$0.02\,$\pm$\,0.03, with a 16--86$^{\rm th}$ per centile range of $\Delta z$/\,$(1+z_{\rm
spec})\,$\,$= -$0.16--0.10. This photometric redshift accuracy is comparable to that found for SMGs in the COSMOS field by \cite{Battisti19}.   Here we provide a brief description of this analysis, full details are given in  \cite{Ikarashi20}.

We used the latest version of {\sc magphys} 
\citep{Battisti19}, which is optimised to fit SEDs of high redshift, star-forming galaxies
and can provide estimates of the redshifts of the sources based on the SED fitting. 
{\sc magphys} employs an energy balance
technique  to combine information from  the attenuation of the stellar emission
in the UV/optical and near-infrared by dust, and the reradiation of
this energy in the far-infrared.  This is a particular advantage for modelling the
photometric redshifts of
very obscured galaxies such as SMGs, where there may be relatively few constraints
available from the optical and near-infrared SED shape due to the the dust obscuration.

To fit to the observed SED galaxy, {\sc magphys} generates a
library of model SEDs for a grid of redshifts for each star-formation history
considered.  The code selects models that best-fit the
multi-wavelength photometry by matching the model SEDs to the data
using a $\chi^2$ test and returns the respective best-fit parameters, most importantly
it provides a median redshift from the    probability distribution
(PDF) from the best-fit models as well as the full PDF of the redshifts.  
We discuss the photometric redshifts derived from our {\sc magphys} analysis in \S 3.5  and 
the other physical properties of the luminous SMGs from AS2COSMOS in \cite{Ikarashi20}.

\section{Results and Discussion}

We start by discussing the basic  properties of the sample as illustrated in Figure~\ref{fig:multimages} and Figure~\ref{fig:multimages_app}.  Most noticeable in these $K_S 3.6\mu{\rm m} 4.5\mu{\rm m}$ images is the typical faintness and the red colours of the majority of the SMGs (even in this combination of near- and mid-infrared filters) compared to the general field population in these fields.   Next, we note several examples where the SMG lies very close to bright and blue galaxies, which are likely foreground (e.g.\ AS2COS\,0011.1, AS2COS\,0001.1,
AS2COS\,0062.1, etc).  These are likely to be examples of gravitationally-lensed systems, although the typical separation of the SMG from the putative lens suggests few of them are examples
of strongly lensed (multiply-imaged) systems with the highest amplifications. Instead the expected lens boosts are likely to be modest: $\sim $\,1.2--3\,$\times$ (see \S3.4 for an example).    Finally, in the fields which show two or more SMGs, there is a visual impression that these preferentially display separations 
 of $\sim$\,2--5$''$, and a more quantitative analysis suggests a strong excess of pairs of sources on scales
of $\lsim$\,3$''$ (corresponding to $\sim$\,20--30\,kpc at typical redshifts for SMGs).  This 
characteristic scale is smaller than the ALMA primary beam and if real could either be a signature of lensing, or  it could be indicating a natural scale for peak activity of interacting galaxies. We will discuss this issue further in \cite{Ikarashi20}.  We note that the rapid increase in the apparent presence of companion SMGs in the fields
of the fainter sources in Figures~\ref{fig:multimages_app} is simply a result of the fact that these SMGs only come to be in our bright single-dish-selected sample through their presence in the field of a second, brighter SMG.

%
%
\begin{figure*}
    \includegraphics[width=0.89\columnwidth, bb=-10pt -10pt 740.81pt 668.81pt]{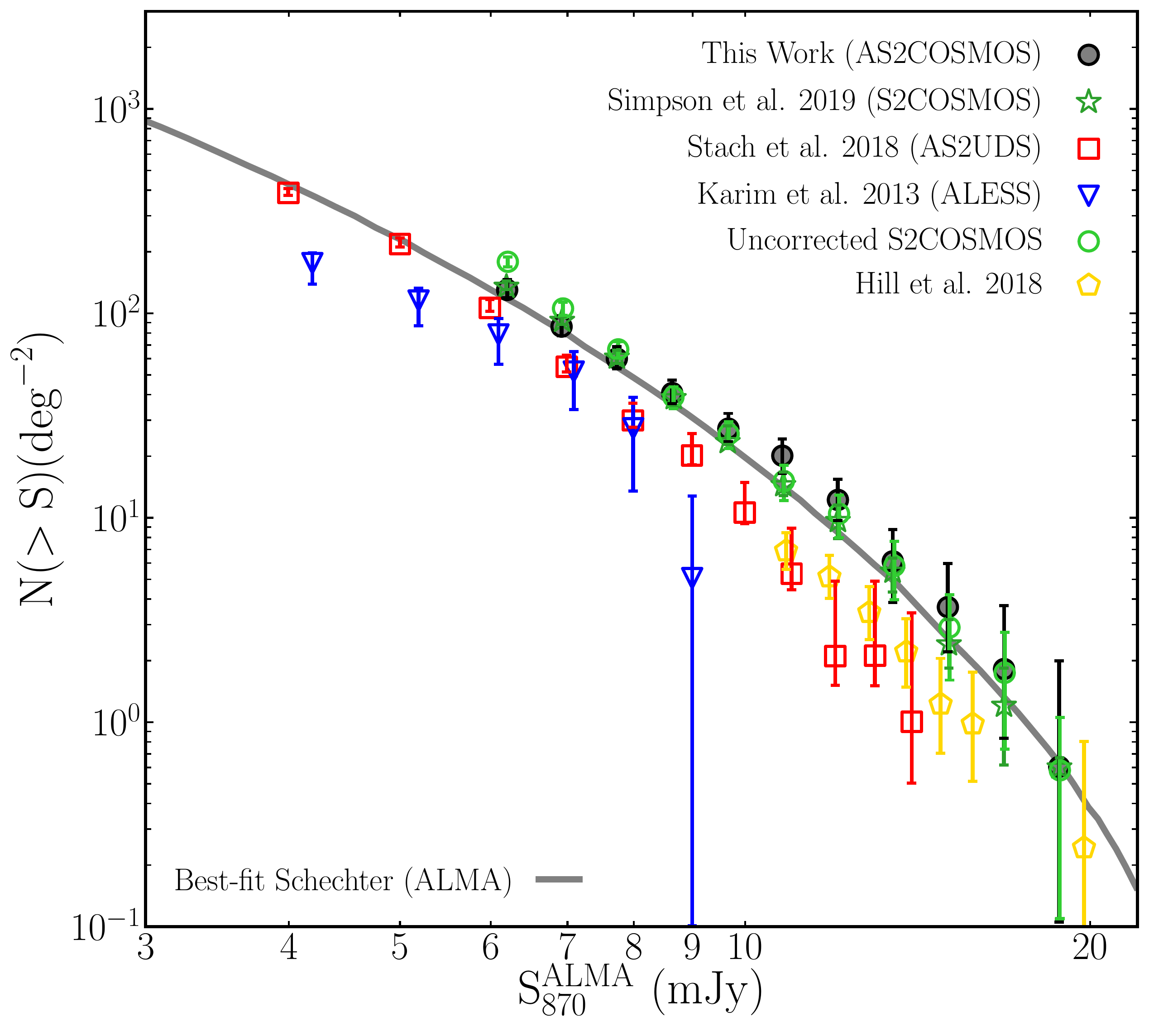}
    \hspace*{0.5cm}
    \includegraphics[width=\columnwidth]{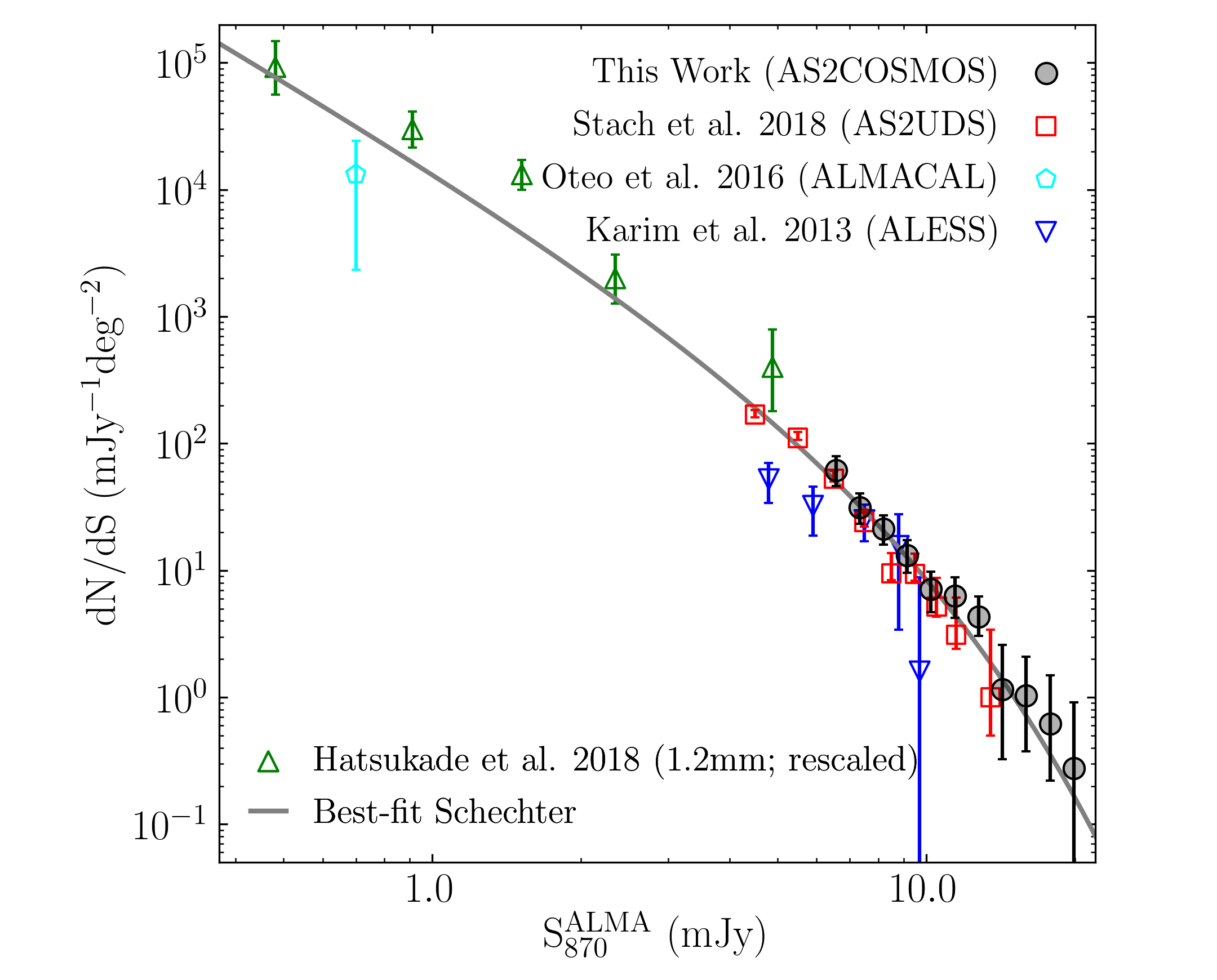}
    \caption{{\it Left:} The cumulative 870-$\mu$m  number counts constructed from the AS2COSMOS pilot survey, compared to those constructed from the parent single-dish SCUBA-2 sample and other interferometric surveys. We find that both the shape and normalisation of the AS2COSMOS counts are in good agreement with those from the S2COSMOS survey after corrections are applied to the latter to account for boosting and blending \citep{Simpson19}, for illustration we also showing the  number counts of sources from S2COSMOS without these corrections.
The AS2COSMOS counts are marginally higher that the boosting/blending-corrected single-dish counts at $\sim$\,10\,mJy, but we stress that any difference is at the $<$\,3\,$\sigma$ significance level. For comparison we also show the counts constructed from prior ALMA studies of LABOCA-selected sources in the ECDFS (ALESS; \citealt{Karim13}), SCUBA-2-selected sources in the UDS (AS2UDS; \citealt{Stach18}) and the SMA-follow-up of bright SCUBA-2 sources from S2CLS \citep{Hill18}. The AS2COSMOS counts are a factor of $\sim$\,1.4--2.0\,$\times$ higher than those constructed from AS2UDS or the study of \citet{Hill18} (which includes the sources in the AS2UDS 
pilot published by \citealt{Simpson15b}), but are consistent within the associated uncertainties at the $<$\,2.3\,$\sigma$ significance level, before considering the effects of cosmic variance (see \S 3.2). {\it Right:} Similar to the left panel but showing the differential 870-$\mu$m  number counts constructed from AS2COSMOS pilot and published ALMA surveys. We also show the counts constructed from deeper, small area surveys with ALMA that were conducted around either calibrator fields (ALMACAL; \citealt{Oteo16}) or as a blank-field mosaic (ASAGAO; \citealt{Hatsukade18}). The differential counts from the various ALMA surveys decline smoothly from $S_{870\mu \rm m}$\,=\,0.4--20\,mJy and are well-described by a single Schechter function. Overall, we highlight that the AS2COSMOS pilot survey detects 108 (39) SMGs at $S_{870\mu \rm m}$\,$>$\,7\,(10)\,mJy, and represents a factor of $\sim$\,2\,$\times$ increase in sample size relative to the largest previous studies. 
    }
    \label{fig:counts}
\end{figure*}

 
\subsection{Number counts}\label{subsec:counts}
\begin{table}
 \centering
 \centerline{\sc Table 2: AS2COSMOS Number Counts}
\vspace{0.1cm}
 {%
 \begin{tabular}{cccc}
 \hline
 \noalign{\smallskip}
 $S_{870}$ & $N(>S_{870})$  & $S_{870}$ & $dN/dS_{870}$  \\ 
 (mJy) & (deg$^{-2}$) & (mJy) & (deg$^{-2}$\,mJy$^{-1}$)  \\ 
\hline \\ [-1.9ex] 
6.2 & 130$^{+13}_{-12}$ & 6.6 & 61.4$^{+20.0}_{-16.0}$ \\ 
6.9 & 85.9$^{+9.5}_{-8.7}$ & 7.3 & 31.2$^{+10.0}_{-8.0}$ \\ 
7.7 & 60.4$^{+7.4}_{-6.7}$ & 8.2 & 21.2$^{+6.0}_{-5.2}$ \\ 
8.6 & 40.9$^{+5.8}_{-5.2}$ & 9.2 & 13.1$^{+4.2}_{-3.6}$ \\ 
9.7 & 27.6$^{+4.8}_{-4.2}$ & 10.2 & ~7.1$^{+2.7}_{-2.4}$ \\ 
10.8 & 20.1$^{+4.1}_{-3.6}$ & 11.4 & ~6.3$^{+2.5}_{-2.0}$ \\ 
12.1 & 12.2$^{+3.3}_{-2.6}$ & 12.8 & ~4.3$^{+1.9}_{-1.3}$ \\ 
13.5 & ~6.1$^{+2.7}_{-2.3}$ & 14.3 & ~1.2$^{+1.4}_{-0.8}$ \\ 
15.1 & ~3.7$^{+2.3}_{-1.5}$ & 15.9 & ~1.0$^{+1.1}_{-0.7}$ \\ 
16.8 & ~1.8$^{+1.9}_{-1.0}$ & 17.8 & ~0.6$^{+0.9}_{-0.4}$ \\ 
18.8 & ~0.6$^{+1.4}_{-0.5}$ & 19.9 & ~0.3$^{+0.6}_{-0.2}$ \\ 
 \hline\hline \\  [0.1ex]  
 \end{tabular}
 \vspace{-0.2cm}
 \begin{flushleft}
 \footnotesize{The cumulative and differential number counts at 870\,$\mu$m constructed from the AS2COSMOS survey of the central 1.6\,deg$^{2}$ of the COSMOS field. The cumulative count bin fluxes are
at the lower limit of the bin and the differential count fluxes refer to the bin centres.}
 \end{flushleft}
 }
 \refstepcounter{table}
 \label{table:obs}
 \end{table}
The number of sub-millimetre emitters as function of flux density, i.e.\ the number counts, is a basic observable property that can provide constraints on models of galaxy formation (e.g.\ \citealt{Baugh05}). The AS2COSMOS sample has a relatively simple selection function (see Figure\,\ref{fig:sampleselection}) and, as such, it is well-suited to constrain the bright-end of the 870-$\mu$m number counts. We determine these AS2COSMOS number counts here and compare our results to previous surveys of luminous SMGs. 

In Figure\,\ref{fig:counts} we present the cumulative and differential number counts derived from the AS2COSMOS source catalogue. The counts are constructed to a lower flux limit of 6.2\,mJy, corrected for sample incompleteness using the completeness curve determined for the sample in \S\,\ref{subsec:completeness} (see Figure\,\ref{fig:sampleselection}), and normalised to the 1.6\,deg$^2$ area of the S2COSMOS {\sc main} survey. The associated uncertainties on the AS2COSMOS counts were estimated by constructing 10$^{4}$ realisations of the AS2COSMOS source catalogue. In each realisation we randomly assigned a flux density to each AS2COSMOS SMG based on its measured value and associated uncertainty and reconstructed the counts. The 16--84$^{\mathrm{th}}$ per centile of the resulting distribution was combined with the expected Poisson uncertainty (\citealt{Gehrels86}) to provide an estimate of the total uncertainty on each bin in the counts.

We find that both the AS2COSMOS cumulative and differential counts follow a smooth, near exponential decline between $S_{870\mu \rm m}$\,=\,6.2 and 20\,mJy. As shown in Figure\,\ref{fig:counts}, the AS2COSMOS cumulative counts are in good agreement with those derived for the S2COSMOS survey -- once allowance has been made in the latter for  the effects of boosting and blending based on a model with a representative functional form for the intrinsic counts.
While a comparison to the {\it raw} uncorrected S2COSMOS counts shows that they  are  $\sim$\,31\,$\pm$\,8 per cent higher
at the survey limit. We note that the ALMA  cumulative counts are marginally {\it higher} than those estimated from the corrected single-dish survey at $\sim$\,10\,mJy, indicating
the limitations of the end-to-end modelling technique (which disregards clustering) to account for blending, 
although any differences are measured are at the $<$\,3\,$\sigma$ significance level (after accounting for the contribution from Poisson noise to the associated uncertainties). The agreement between the ALMA and the SCUBA-2 counts is consistent with our earlier result that the brightest SMG in each AS2COSMOS map accounts for, on average, all of the deboosted and deblended flux density of the targeted SCUBA-2 source (see Figure\,\ref{fig:flux}) -- showing that those statistical corrections are reliable on average, if the approximate form of the counts is already known. 

Previous interferometric follow-up observations of single-dish-identified sub-millimetre sources have  reported similar reductions in the normalisation of the interferometric counts, relative to the parent single-dish sample (e.g.\ \citealt{Karim13,Simpson15a,Stach18,Hill18}). For example, \citet{Stach18} present the number counts derived from the AS2UDS survey of 714 SCUBA-2 sources in the UDS field, and report a 28\,\,$\pm$\,2\,per cent (41\,\,$\pm$\,8\,per cent) reduction in the counts at $>$\,4\,mJy ($>$\,7\,mJy), relative to the parent single-dish sample.   This is broadly consistent with the reduction we find when comparing the
uncorrected S2COSMOS counts to those derived here, as expected  given our  result below that the parent samples for both surveys suffer a comparable level of source blending (see \S\,\ref{subsec:mult}).  

Figure\,\ref{fig:counts} also shows that our ALMA-derived counts in the COSMOS field lie a factor of  1.4--2.0\,$\times$ higher than those from AS2UDS or the SMA study of bright S2CLS sources by \citep{Hill18}.  This difference  corresponds to a  formal significance of $\sim$\,2.3\,$\sigma$ at the limit of AS2COSMOS. But as we show in \S3.2, these studies are broadly consistent when allowance is made for the cosmic variance in the counts derived
from similar sized areas drawn from simulations created using the {\sc galform} semi-analytic galaxy formation model.

To provide a simple parameterisation of the sub-millimetre number counts we now determine the best-fit model to both the AS2COSMOS counts and prior estimates of the sub-millimetre counts based on sensitive ALMA observations in the literature. At the bright-end of the counts ($\gsim$\,4\,mJy), we include in our analysis the estimates of 870-$\mu$m counts from the AS2UDS and ALESS surveys \citep{Karim13,Stach18}, and extend our analysis to fainter fluxes by including the estimate of the 870-$\mu$m counts from the ALMACAL survey \citep{Oteo16} and the 1.2-mm counts from the ASAGAO survey (\citealt{Hatsukade18}; see also \citealt{Dunlop16,Franco18}). The 1.2-mm counts are converted to 870\,$\mu$m assuming a modified blackbody with $\beta=$\,1.8 and dust temperature of 32\,K, at a redshift of $z=$\,2.6 (e.g.\ \citealt{Dudzeviciute20}; $S_{870\mu \rm m}$\,/\,$S_{1200}=$\,2.7). We assume that the sub-millimetre counts follow a simple Schechter function (Equation~\ref{eqn:Schechter}) of the form:

\begin{equation}\label{eqn:Schechter}
\frac{dN}{dS} = \frac{N_{0}}{S_0} \left( \frac{S}{S_0}\right)^{-\gamma} \mathrm{exp}\left( \frac{-S}{S_0} \right).
\end{equation}

 \noindent and determine best-fit parameters of $N_{0}$\,=\,2770$^{+1560}_{-650}$\,deg$^{-2}$, $S_{0}$\,=\,$4.2^{+0.5}_{-0.8}$\,mJy, and $\gamma$\,=\,2.3$^{+0.1}_{-0.3}$. As can be seen in Figure\,\ref{fig:counts} the best-fit parameterisation provides a reasonable description of the observed counts (reduced $\chi^{2}=$\,1.5) at $S_{870\mu \rm m}=$\,0.4--20\,mJy, although we note that the faint-end of the counts ($\lsim$\,4\,mJy) is constrained by a modest number of source ($\sim$\,40--50) and that this is reflected in the significant associated uncertainties on our best-fit model parameters.

%
%
\begin{figure}
    \includegraphics[width=\columnwidth]{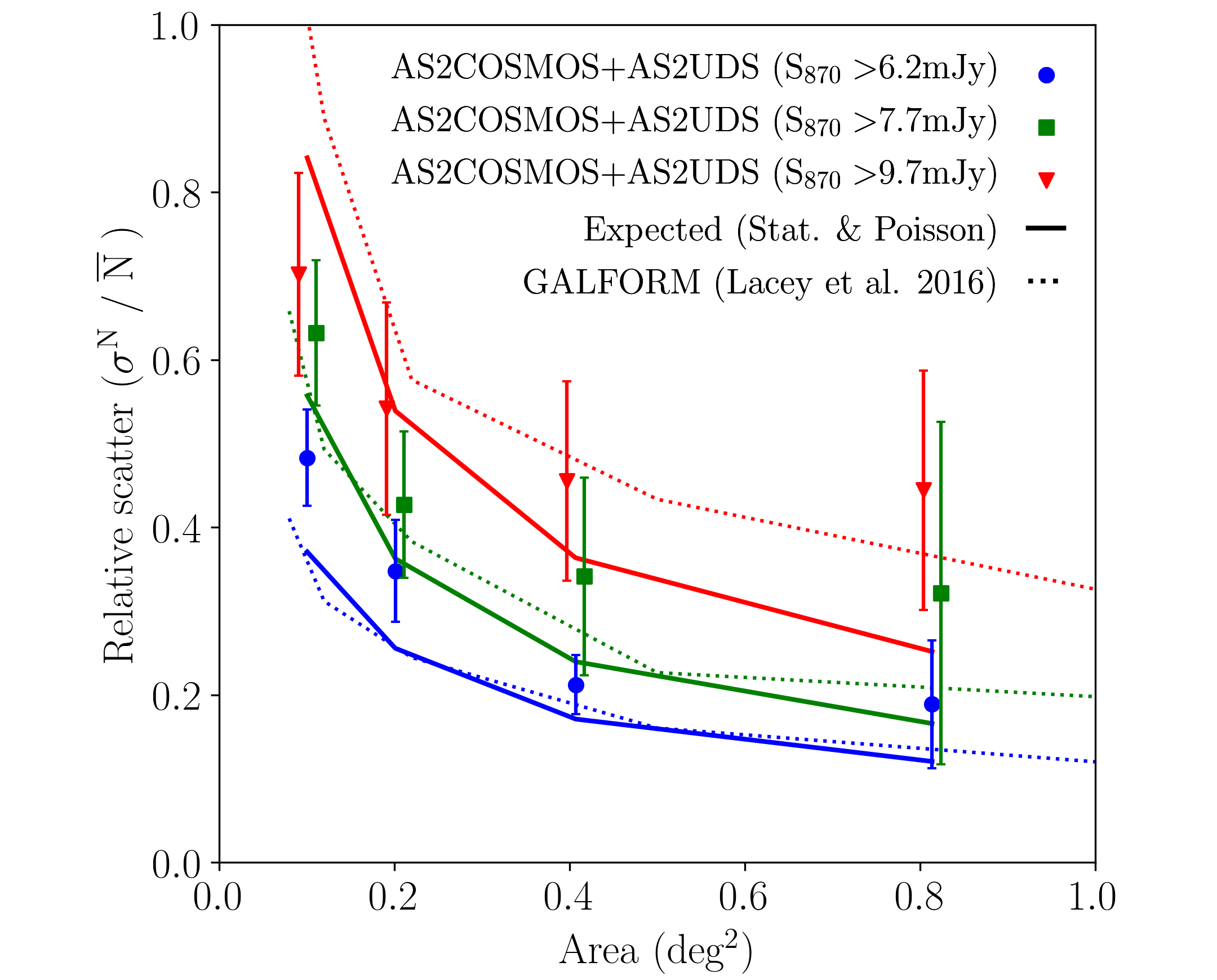}
    \caption{ The fractional scatter in the cumulative 870-$\mu$m number counts as a function of survey area and flux limit: $\sigma^N/\bar{N}$. Number counts were constructed from sub-areas of the AS2COSMOS and AS2UDS surveys and the scatter in the results, $\sigma^N$, is normalised to the sample mean, $\bar{N}$. The expected scatter in the observed counts is represented by a solid line and reflects the contribution from both statistical (e.g.\ flux uncertainties) and Poissonian uncertainties. We find an enhancement in the relative scatter of the cumulative counts that can be attributed to cosmic variance of $\sim$\,30\,per cent in the $S_{870\mu \rm m}>$\,6.2\,mJy population in survey areas of $\lsim$\,0.2\,deg$^{2}$, although we  caution that the significance of the result is modest. For comparison, we show the total scatter estimated from the {\sc galform} semi-analytic model of galaxy formation (dotted line). The predictions of the {\sc galform} model are broadly consistent with the results presented here, and suggests that
the observed $\gsim$\,1.4\,$\times$ difference between the cumulative source counts in the AS2COSMOS and AS2UDS surveys could
be simply due to cosmic variance.
     }
    \label{fig:countsvar}
\end{figure}

\subsection{Cosmic Variance}
Since the discovery of the SMG population it has been speculated that these intensely star-forming systems may be the progenitors of local spheroidal galaxies (e.g.\ \citealt{Lilly99,Blain04,Swinbank06,Tacconi08,Swinbank10,Simpson14,Dudzeviciute20}). Under the $\Lambda$CDM paradigm, SMGs would thus represent a  biased tracer of the underlying mass distribution of the Universe (e.g.\ \citealt{Hickox09}), which we would expect to manifest as excess field-to-field variance in the integrated 870-$\mu$m number counts \citep{Scott12}.  

Exploiting the AS2COSMOS and AS2UDS surveys we can investigate whether the interferometrically-identified SMG population (so unaffected by blending) shows evidence for cosmic variance as a function of both 870\,$\mu$m flux density and survey area. The AS2COSMOS and AS2UDS surveys are homogenous and, taken together, provide a catalogue of 223 bright SMGs ($S_{870\mu \rm m}$\,$>$\,6.2\,mJy) selected over a survey area of 2.6\,deg$^2$, corresponding to survey volume of 0.12\,Gpc$^{3}$ if we assume a maximal redshift range for the SMG population of $z$\,$\sim$\,1--5 (see \S\,\ref{subsec:redshifts}, but also \citealt{Simpson14,Strandet16,Dudzeviciute20}). 

To investigate the effect of cosmic variance on the bright 870-$\mu$m ALMA number counts we first sub-divide the AS2COSMOS and AS2UDS surveys into 26 regions each with an area of $\sim$\,0.1\,deg$^2$. These sub-regions were then combined to provide a sample of representative surveys over 0.1--0.8\,deg$^2$. For each sub-field we derived the completeness-corrected, integrated counts and estimated the total variance in the resulting distribution, normalised to the sample mean: $\sigma^N/\bar{N}$. The total variance in the counts is comprised of contributions from cosmic variance, Poisson errors, and statistical uncertainties (e.g.\ flux density estimates). To estimate the statistical uncertainty on the distribution of counts we use a set Monte Carlo simulations, comprising 10$^3$ realisations of the integrated counts for each sub-field. The expected Poisson uncertainty is estimated following \citet{Gehrels86}.

In Figure\,\ref{fig:countsvar} we show the total variance in the cumulative number counts as a function of survey area, as well as in three  bins of the AS2COSMOS 870-$\mu$m counts ($S_{870\mu \rm m}$\,$>$\,6.2, 7.7, and 9.7\,mJy). As expected, the total field-to-field variance in the 870-$\mu$m counts increases in smaller areas and at higher  flux densities. At $S_{870\mu \rm m}$\,$>$\,6.2\,mJy we estimate that the total normalised variance in the counts decreases from 48\,$\pm$\,6\,per cent over survey regions of 0.1\,deg$^2$ to $\sim$\,20\,per cent at 0.4--0.8\,deg$^2$ (21\,$\pm$\,4\,per cent at 0.4\,deg$^2$). To isolate any potential contribution from cosmic variance, we subtract the estimated statistical and Poissonian uncertainties from the total variance, as a function of survey area and flux density (see Figure\,\ref{fig:countsvar}). We estimate cosmic variance of $\sim$\,30 and $\sim$\,20\,per cent in the $>$\,6.2\,mJy population for survey areas of 0.1 and 0.2\,deg$^2$, respectively, but  caution that the excess variance is only significant at the $\lsim$\,2\,$\sigma$ level (1.9\,$\sigma$ and 1.6\,$\sigma$, respectively). If we consider larger survey areas, or brighter flux limits, then we determine that the total variance is systematically elevated, relative to that expected from Poisson uncertainties and other errors, but  again this is at the $\ll$\,1\,$\sigma$ significance level. 

To assess how the variance we measure compares to that expected from theoretical models we use results from the {\sc galform} semi-analytic model of galaxy formation  \citep{Lacey16}. We construct a 20\,deg$^2$ of model sub-millimetre sky using  {\sc galform}, in five distinct light-cones, and estimate the normalised variance in the integrated number of simulated sub-millimetre galaxies in sub-regions spanning 0.1--1.0\,deg$^2$. The {\sc galform} simulations do not include any statistical uncertainties and as such, to ensure consistency with the observational results, we add our estimate for the statistical variance on the AS2COSMOS\,$+$\,AS2UDS sample to the measured variance in the simulation. As shown in Figure\,\ref{fig:countsvar}, the predictions from {\sc galform} are broadly inline with our observational results, although we note that the variance in the simulation lies systematically below our observational result for the $S_{870\mu \rm m}$\,$>$\,6.2\,mJy population on 0.1--0.2\,deg$^2$ scales (this may be explained by the number of simulated sources being higher than observed, so the relative Poisson contribution is lower).   We also note that  the observed AS2UDS counts are a factor of $\sim$\,1.4--2.0\,$\times$ lower than those from AS2COSMOS (Figure\,\ref{fig:counts}), which corresponds to a  formal significance of $\sim$\,2.3\,$\sigma$.  However,  the ALMA counts in the two fields are broadly consistent with the scatter between degree-sized fields predicted by {\sc galform}.

We highlight that these empirical limits on the cosmic variance in the counts of SMGs in $\sim$\,0.1\,deg$^2$ areas have implications for the searches for overdensities of such sources which rely on identifying the excess in the projected surface density of sources, unless care is taken to assess the significance of above-Poisson variance
in the number counts \citep[e.g.,][]{Dannerbauer14,Casey16,Harikane19}.

%
%
\begin{figure*}
    \includegraphics[width=\columnwidth]{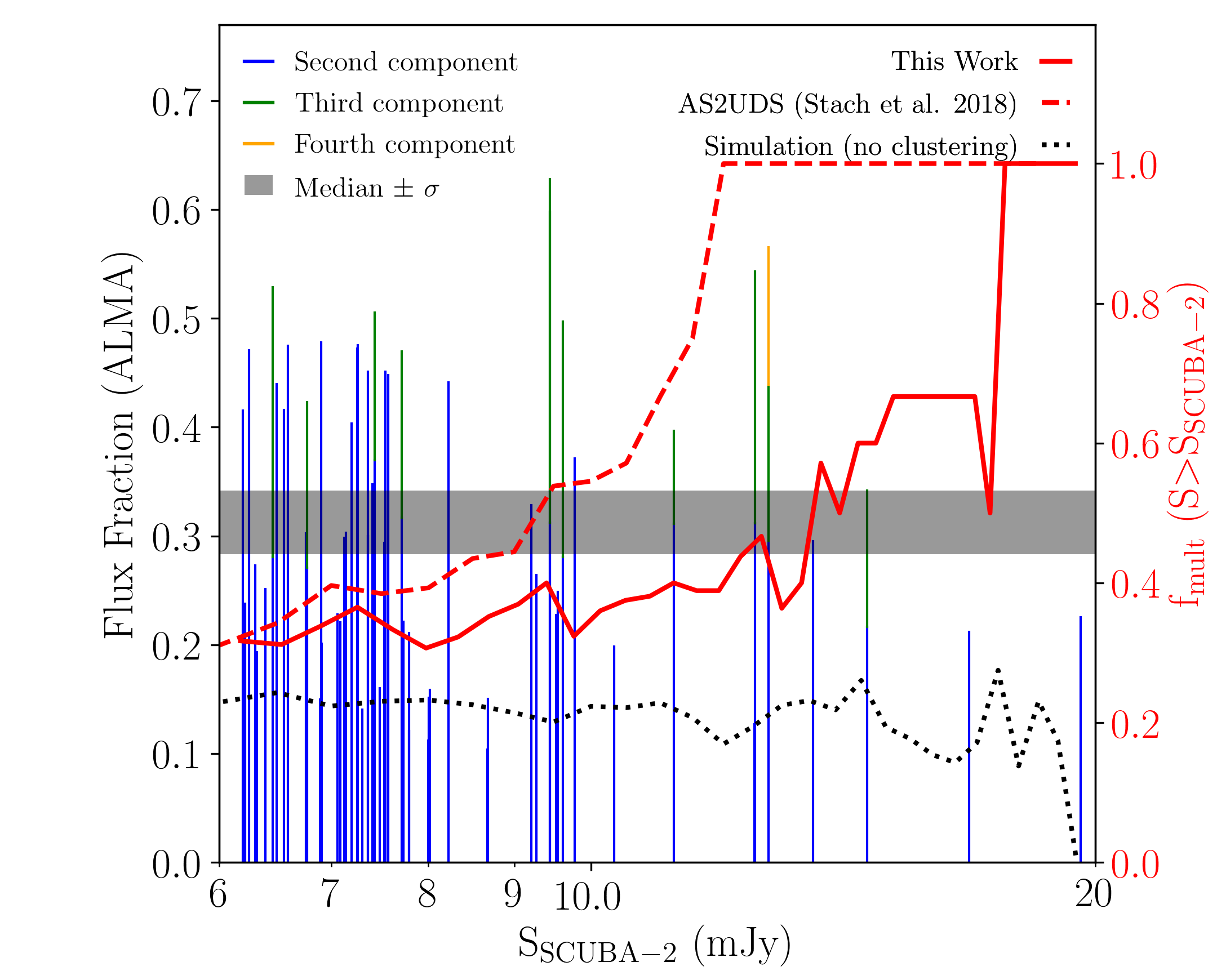}
  	\hfill
    \includegraphics[width=\columnwidth]{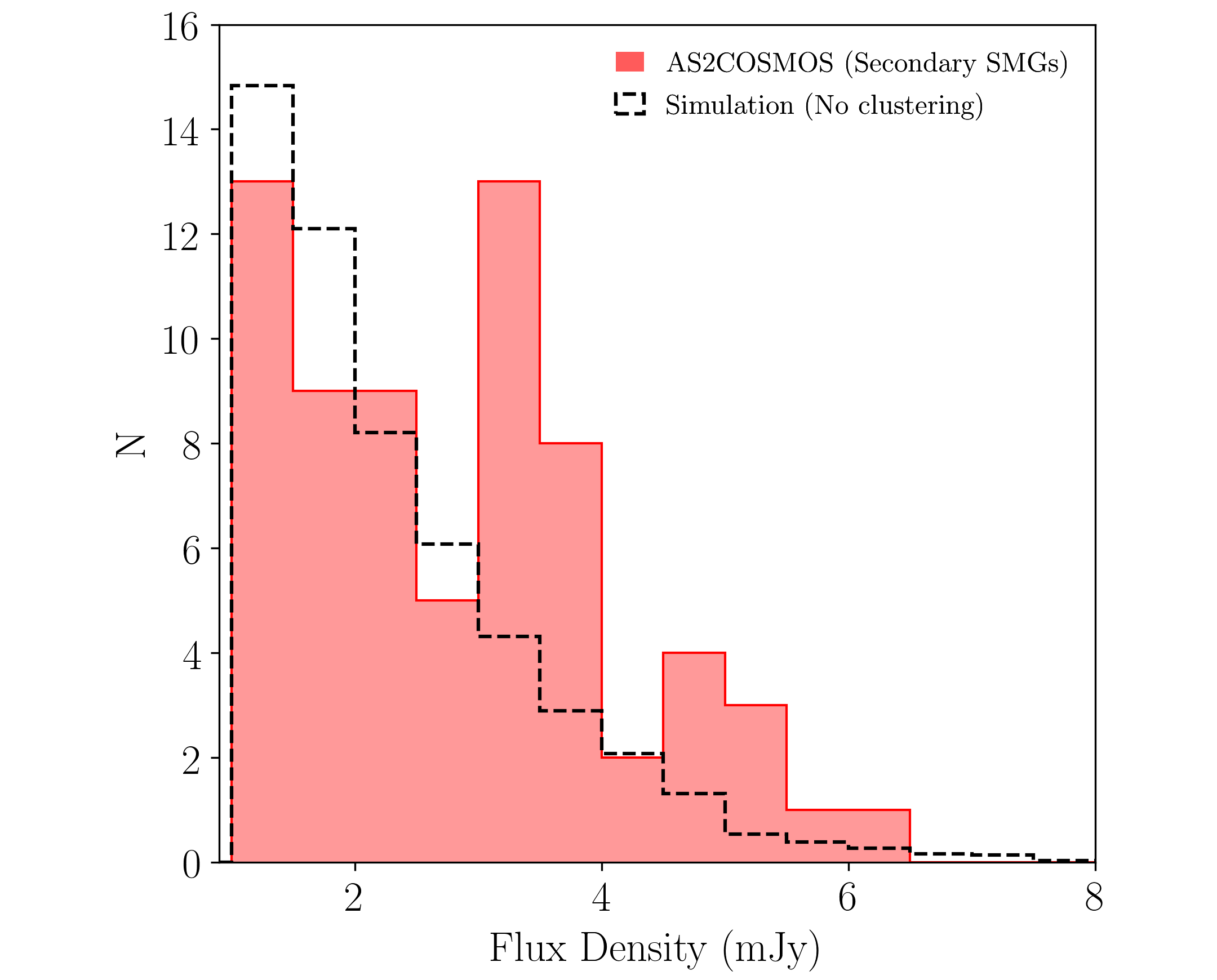}
    \caption{{\it Left:} The fraction of the integrated ALMA flux density that is contained in secondary components ($S_{870\mu \rm m}$\,$>$\,1\,mJy) in each AS2COSMOS map, as a function of the single-dish flux density of the targeted S2COSMOS source. Secondary AS2COSMOS SMGs contribute, on average, 30$^{+3}_{-2}$\,per cent (shaded region) of the integrated ALMA flux density with no significant dependance on SCUBA-2 flux density, in good agreement with results from AS2UDS for the same range in flux density (30\,$\pm$\,1\,per cent; \citealt{Stach18}). The fraction of AS2COSMOS maps that contain multiple SMGs (dashed line) shows a weak dependence on single-dish flux density, in broad agreement with the AS2UDS survey (dot-dash line; \citealt{Stach18}). The level of multiplicity in the AS2COSMOS and AS2UDS samples is elevated relative to the expectations of an unclustered population (dotted line), which suggest that $\sim$\,30\,per cent of the AS2COSMOS ``multiple'' maps contain physically-associated SMGs. {\it Right:} The flux density distribution of the 76 secondary SMGs ($S_{870\mu \rm m}$\,$>$\,1\,mJy) that are detected within the primary beam of each AS2COSMOS map. For comparison, we show the expected flux distribution distribution of secondary components from our simulation of the A\,/\,S2COSMOS surveys, which assume an unclustered population of sub-millimetre sources. We find that the number density of secondary AS2COSMOS SMGs with flux densities $<$\,3\,mJy is broadly consistent with the results of our simulation, indicating that these faint SMGs are typically line-of-sight associations to the primary SMG in each ALMA map. However, we find a clear excess in the number density of AS2COSMOS secondaries brighter than $\gs$\,3\,mJy, relative to the simulation, and estimate that 62\,$\pm$\,7\,per cent of these brighter components are physically-associated with the primary SMG in their respective maps. 
    }
    \label{fig:mult}
\end{figure*}

\subsection{Multiplicity}
\label{subsec:mult}
Using our  catalogue of AS2COSMOS sources we now assess the level of multiplicity in the parent S2COSMOS sample. We follow the convention adopted in prior studies (e.g.\ \citealt{Simpson15a,Cowie18,Stach18}) and define a single-dish S2COSMOS source as a ``multiple'' if two-or-more SMGs with flux densities $S_{870\mu \rm m}\ge$\,1\,mJy are identified within the primary beam of the corresponding ALMA map (i.e.\ within 8.7$''$ of the SCUBA-2 position).
We find  one single-dish source which breaks up into four SMG counterparts, eleven which are blends of three SMGs and a further 51 with two SMGs counterparts (Figures~\ref{fig:fieldplan}, \ref{fig:multimages} \& \ref{fig:multimages_app}).  The highest multiplicity source, S2COSMOS\,0003 (Fig.~\ref{fig:fieldplan}),
has been previously discussed by \cite{Wang16} who have shown that the four components all lie in 
a single structure at $z=$\,2.50.  We discuss a similar association of dusty star-forming galaxies associated
with the highest significance S2COSMOS source, S2COSMOS\,0001, at $z=$\,4.63 in \S 3.4.

In total, we find that 63 of the 182 AS2COSMOS maps contain two or more SMGs with $S_{870\mu \rm m}>$\,1\,mJy, corresponding to a multiplicity fraction for the sample of 34\,$\pm$\,2\,per cent.   These secondary SMGs contribute a median of 30$^{+4}_{-2}$\,per cent of the integrated ALMA flux density of all sources in each ALMA map, and we find no evidence that this fraction depends on single-dish flux density in the flux range we probe (Figure~\ref{fig:mult}).
The level of multiplicity in the AS2COSMOS sample is significantly higher than the 11\,$\pm$\,1\,per cent determined by \citet{Stach18} for the typically fainter AS2UDS sample, or the 13\,$\pm$\,6\,per cent found by \citet{Cowie18} in their Super-GOODS survey\,\footnote{Source multiplicity is sensitive to the the beam size of the parent single-dish observations and the depth and resolution of the follow-up interferometric imaging. As such, we choose to focus our comparison on prior studies that obtained ALMA follow-up observations of SCUBA-2-identified sources but note that comparable results have been obtained in studies of SMGs with other facilities (e.g.\ \citealt{Karim13, Brisbin17, Hill18})}. However, the multiplicity of SCUBA-2 sources has been shown to correlate with their single-dish flux density (\citealt{Stach18}), and both the AS2UDS and Super-GOODS samples probe to fainter fluxes that the sources considered here ($S_{870\mu \rm m}$\,$>$\,4\,mJy and $S_{870\mu \rm m}$\,$>$\,2.2\,mJy, respectively). To provide a fair comparison between the AS2COSMOS and AS2UDS surveys we estimate the multiplicity of the 88 AS2UDS sources with single-dish flux densities brighter than the selection limit for our pilot AS2COSMOS sample (i.e.\ $S_{870\mu \rm m}$\,$>$\,6.2\,mJy). We do not consider the Super-GOODS sample presented by \citealt{Cowie18} as it contains only four SCUBA-2 sources with flux densities above the S2COSMOS selection limit. For the subset of 88 AS2UDS sources brighter than $S_{870\mu \rm m}=$\,6.2\,mJy we determine a multiplicity rate of 33\,$\pm$\,5\,per cent (and a median fractional flux in secondaries of 30\,$\pm$\,1 per cent). This is in very good agreement with our results for AS2COSMOS and confirms the overall agreement between the two surveys.

Our analysis highlights an apparent change in the multiplicity rate of SCUBA-2 sources between $S_{850\mu\rm m}$\,$\sim$\,4 and 6\,mJy \citep{Stach18}. Interestingly, there is no evidence in the AS2COSMOS sample for a change in the multiplicity of SCUBA-2 sources at $>$\,6\,mJy (see Figure\,\ref{fig:mult}). To investigate this further, we combine the AS2COSMOS and AS2UDS samples and repeat our analysis on this 50\,per cent larger sample. For the combined sample, we again find no trend in the rate of multiplicity with single-dish flux density between $S_{850\mu\rm m}$\,=\,6--12\,mJy (30\,$\pm$\,2\,per cent). However, there is a statistically-significant increase in the multiplicity rate to 53\,$\pm$\,8\,per cent for sources at $>$\,12\,mJy. Thus, while the frequency of source blending is broadly uniform across the AS2COSMOS sample we note that it does tend to increase for the most luminous of 850-$\mu$m sources.

%
%
\begin{figure*}
    \includegraphics[width=1.6\columnwidth]{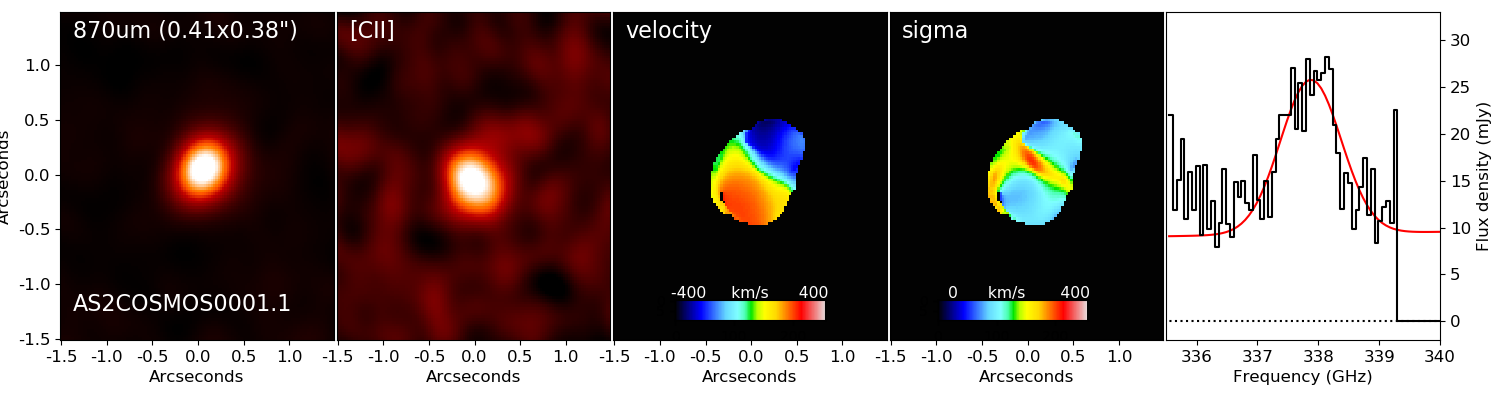}
    \vfill
    \includegraphics[width=1.6\columnwidth]{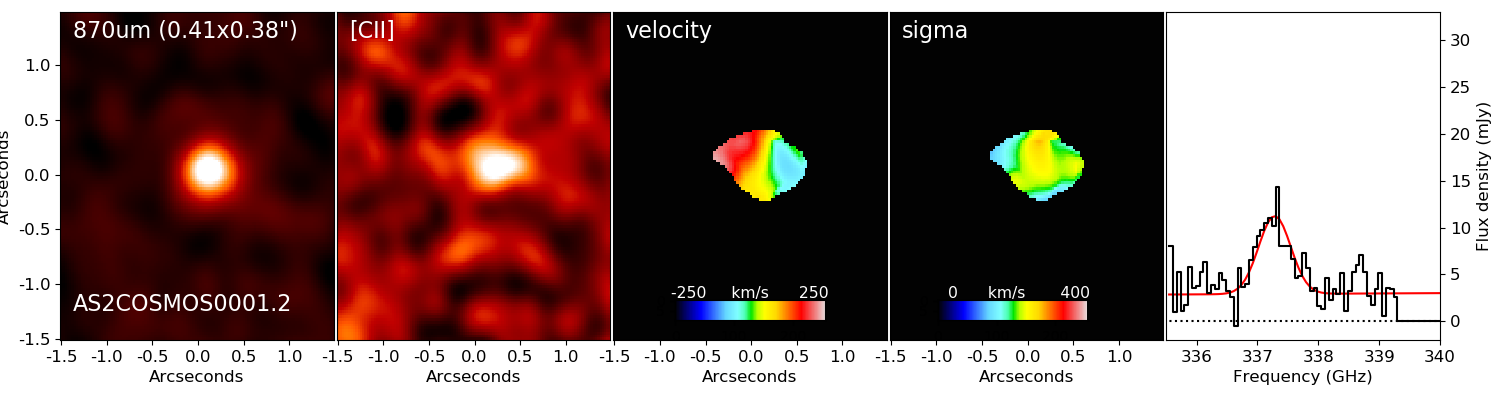}

    \caption{Spatially-resolved dust and line emission from AS2COS\,0001.1\,\&\,0001.2. From left-to-right: (1) observed 870-$\mu$m continuum dust emission; (2) observed, continuum-subtracted [C{\sc ii}] line emission;  (3) velocity profile derived from the [C{\sc ii}] kinematics and (4) velocity dispersion, as estimated from the best-fit Gaussian model to the emission in each spaxel; (5) the spatially-integrated [C{\sc ii}] spectrum for each source, along with a Gaussian fit to the lines (the gap between the two pairs of basebands means there is no data above $\sim$\,339.2\,GHz). The redshifts derived from the  [C{\sc ii}] emission for the two SMGs are
$z$\,=\,4.624 and 4.635. We find that the kinematics of both sources show evidence for a clear velocity gradient and a centrally-concentrated velocity-dispersion, indicative of disk-like rotation in the [C{\sc ii}] emitting gas. The sources have an on-sky separation of 3.1$''$ ($\sim$\,20\,kpc) and velocity-seperation of 590\,$\pm$\,40\,km\,s$^{-1}$, suggesting that they are physically-associated within the same dark matter halo.
    }
    \label{fig:cii}
\end{figure*}

It is important to note that the AS2COSMOS sample is very incomplete for sources with flux densities as faint as 1\,mJy. Hence, our estimate of the multiplicity of SCUBA-2 sources is undoubtably a lower limit. However, we also stress that assuming the best-fit parameterisation of the sub-millimetre number counts (see \S\,\ref{subsec:counts}) we should expected a surface density approximately one SMG brighter than $>$\,1\,mJy per 4--5 ALMA primary beams. This indicates that our adopted definition for a ``multiple'' is beginning to approach the background population.

To quantify the fraction of AS2COSMOS multiples that likely arise from line-of-sight associations we use the suite of A\,/\,S2COSMOS simulations that were presented in the previous sections.
From this ALMA-SCUBA-2 simulation we estimate that 23\,per cent of simulated SCUBA-2 sources with flux densities $>$\,6.2\,mJy would be classed as multiples in a follow-up survey equivalent to AS2COSMOS. The input model for the simulation does not include clustering and, as such, all of the multiples arise due to chance associations along the line-of-sight, rather than physically-associated systems. The fraction of SCUBA-2 sources with multiple ALMA-detected counterparts in AS2COSMOS is 34\,$\pm$\,2\,per cent, which is significantly higher than the prediction of our simulation. Taken together, these results indicate that $\sim$\,30\,per cent of the AS2COSMOS multiple maps contain SMGs that are physically-associated, a rate that is in good agreement with prior studies of a handful of spectroscopically-identified pairs (\citealt{Wardlow18,Hayward18}) or statistical analysis using photometric redshifts (e.g.\ \citealt{Stach18}).

Finally, we consider whether the fraction of multiples that are physically-associated is correlated with the flux density of the secondary SMGs. In Figure\,\ref{fig:mult} we show the 870\,$\mu$m flux density distribution of the 76 secondary AS2COSMOS SMGs that are detected within the primary beam of the AS2COSMOS maps. We find that number of secondary sources rises slowly with flux above our
limit of $S_{870\mu \rm m}=$\,6.2\,mJy.  Figure\,\ref{fig:mult}  also shows the expected flux distribution of these secondary components, as estimated from our end-to-end simulations of the A\,/\,S2COSMOS simulation. As can be seen in Figure\,\ref{fig:mult}, the observed population of secondary AS2COSMOS SMGs with fluxes $\le$\,3\,mJy is broadly consistent with the results of the simulation; the AS2COSMOS sample contains 41 secondaries with fluxes densities of $S_{870\mu \rm m}$\,=\,1--3\,mJy, which agrees precisely with the expected rate of 41 sources from the simulation. However, when we consider secondary sources brighter than $S_{870\mu \rm m}=$\,3\,mJy we  find clear evidence of an excess of secondary  SMGs in AS2COSMOS relative to the simulation.  There are 35 AS2COSMOS secondary SMGs at $S_{870\mu \rm m}$\,$>$\,3\,mJy and we estimate that this is a factor of 2.6\,$\pm$\,0.5\,$\times$ higher than expected from an unclustered population. 

Any multiplicity in our end-to-end simulation of the AS2COSMOS  pilot survey arises due to line-of-sight projections with the primary SMG in each simulated ALMA map. As such, our results indicate that the observed population of ``faint'' AS2COSMOS secondaries ($S_{870\mu \rm m}$\,$<$\,3\,mJy) is overwhelmingly dominated by sources
seen in projection along the line-of-sight  to the primary SMG in each ALMA map. However, where an AS2COSMOS secondary is detected with a flux density of $>$\,3\,mJy we estimate that there is a 62\,$\pm$\,7\,per cent chance that it is physically-associated with the brighest SMG in the map, a significantly higher rate of association than we estimated for the overall sample ($\sim$\,30\,per cent).   Unfortunately, current observational constraints on the
relative mix of projected and associated companions in blended SMG maps are weak and so cannot yet
provide a conclusive test of these estimates \citep[e.g.][]{Wardlow18,Hayward18,Stach18}.  Nevertheless, we have a serendipitous detection of an example of one of these physically associated $S_{870\mu \rm m}$\,$\gs$\,3\,mJy secondary SMGs which we discuss next.

\subsection{AS2COS\,0001.1 \& 0001.2}
The ALMA observations of the S2COSMOS sample were intended to  yield detections of the continuum dust emission
from these soruces. However, the data is also sensitive to any line emission that serendipitously falls within the available 7.5-GHz bandwidth (e.g.\ \citealt{Swinbank12}). 
As described by  \cite{Mitsuhashi20} we searched the ALMA  data cubes for strong emission lines
at the position of each AS2COSMOS source and identified bright line emission from five sources, including both
counterparts to the bright SCUBA-2 source S2COSMOS\,0001\footnote{Also known as AzTEC\,2 \citep{Scott08}}:  AS2COS\,0001.1\,\&\,0001.2, which we discuss in more detail here.  
We note that these sources have also been discussed in a very recent
paper by \cite{JimenezAndrade20}, where they are named AzTEC2-A and AzTEC2-B respectively, and we compare our results to those from their analysis in the following.   We  note that the $\sim$\,2 per cent detection rate of line emitters in the AS2COSMOS data cubes
is comparable to that found in previous SMG studies \citep{Swinbank12,Cooke18}.  A detailed discussion of all potential line emitters in the AS2COSMOS pilot survey is presented in \citep{Mitsuhashi20}.

AS2COS\,0001.1\,\&\,0001.2 were identified in an archival ALMA observation of S2COSMOS\,0001, the highest significance source in the S2COSMOS survey (SNR\,=\,28, $S_{850\mu\rm m}$\,=,\,16.8\,mJy).  The  ALMA-detected SMGs have  870-$\mu$m flux densities of 13.5\,$\pm$\,0.3 and 3.6\,$\pm$\,0.2\,mJy --  where
these fluxes are based on the line-free continuum in the ALMA cubes.   As discussed by  \cite{Mitsuhashi20}
and \cite{JimenezAndrade20}, the ALMA observations of both these sources  show strong lines at $\sim$\,337\,GHz with
apparently high equivalent widths (see Figure~\ref{fig:cii}).

The $^2$P$_{3\,/\,2}$\,$\to$\,$^2$P$_{1\,/\,2}$ fine structure line of atomic carbon (C$^+$) at 157.7\,$\mu$m, hereafter [C{\sc ii}], is typically the strongest far-infrared emission line in the spectra of star-forming galaxies (e.g.\ \citealt{Brauher08,Santos13}). The [C{\sc ii}] emission can comprise 2\,per cent of the total bolometric luminosity of a source, and is typically an order of magnitude brighter than other atomic or molecular emission (e.g.\ [N{\sc ii}]\,122\,$\mu$m, [O{\sc i}]\,145\,$\mu$m, [N{\sc ii}]\,205\,$\mu$m, or mid-$J$ $^{12}$\,CO). At the depth of our observations, [C{\sc ii}] is the most likely identification for the line emission from AS2COS\,0001.1\,\&\,0001.2, placing these sources at a redshift of $z$\,=\,4.624\,$\pm$\,0.001 and 4.635\,$\pm$\,0.001, respectively. This identification has now been unambiguously confirmed by the observations of $^{12}$CO(5--4) emission in these sources  by \cite{JimenezAndrade20} and our
redshift measurements agree within the errors with those from \citealt[][]{JimenezAndrade20}.
These two SMGs lie near to a foreground $z=$\,0.34 galaxy (see Fig.~\ref{fig:multimages})
which results in 
amplifications of $\mu_{1.1}\sim$\,1.5 and  $\mu_{1.2}\sim$\,1.35  
as estimated \cite{JimenezAndrade20}.  We have not corrected for this amplification in the following.

%
%
\begin{figure*}
    \includegraphics[width=\columnwidth]{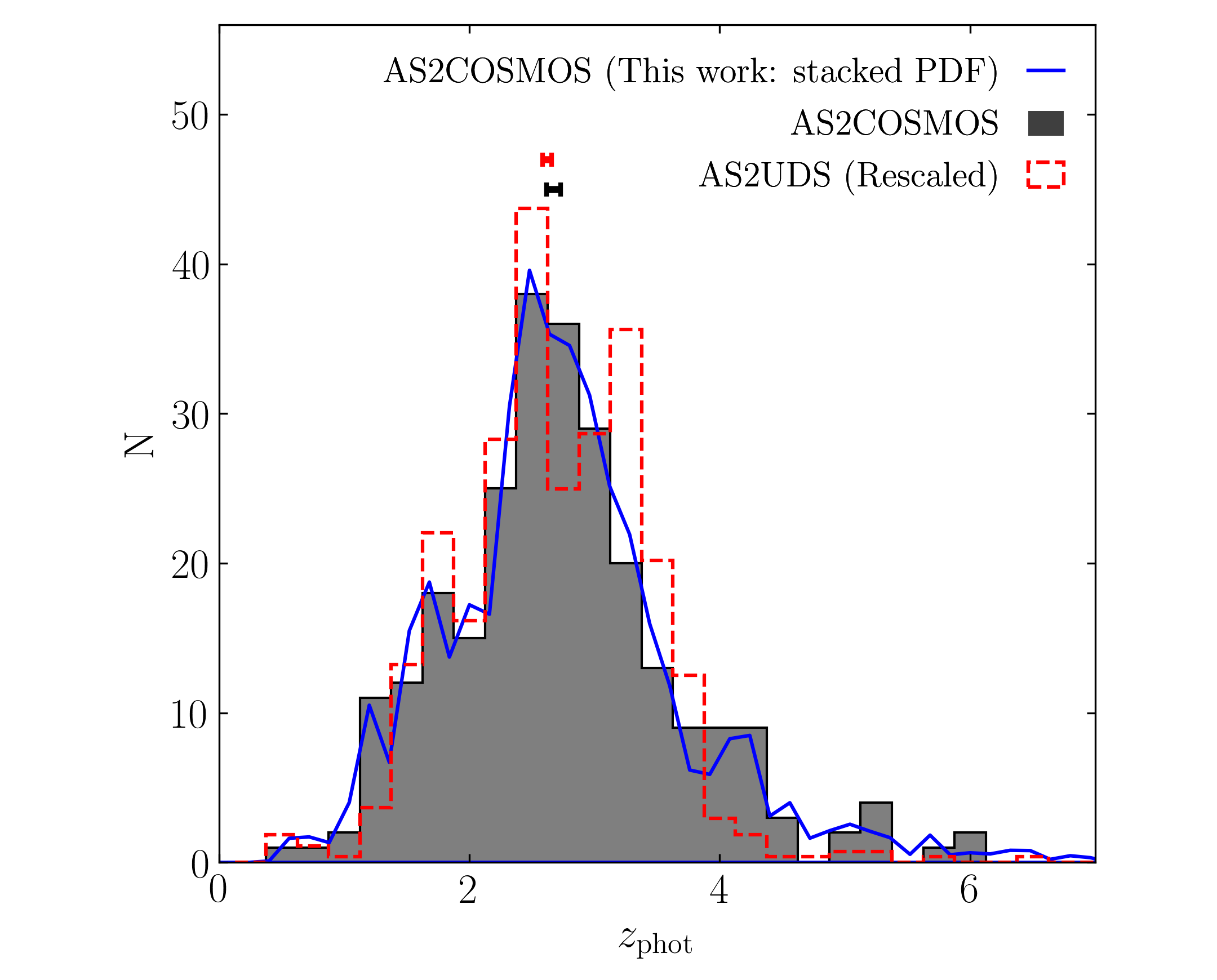}
    \hfill
    \includegraphics[width=\columnwidth]{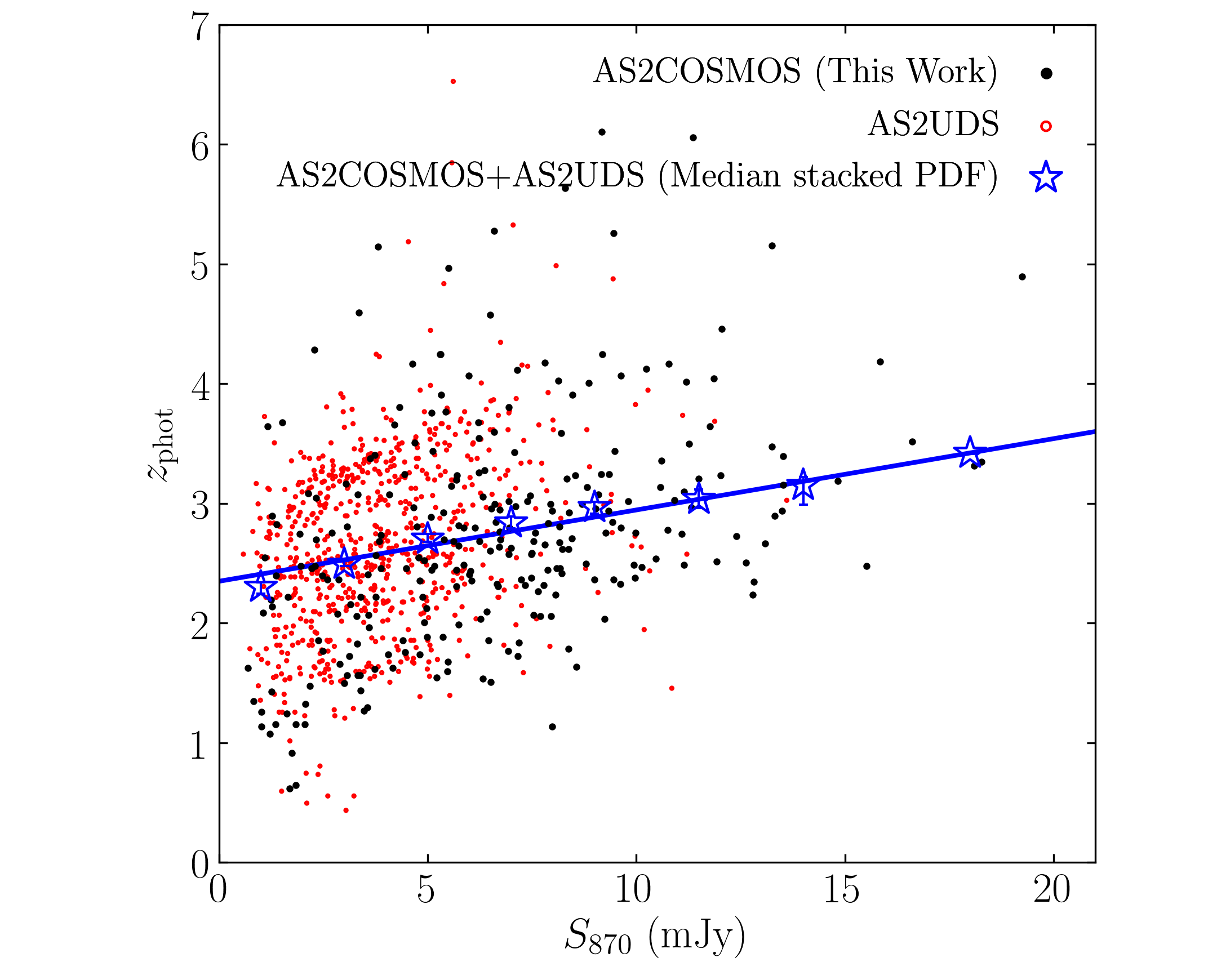}

    \caption{{\it Left:} The  photometric redshift distribution of the AS2COSMOS SMGs, as determined from our {\sc magphys} analysis of their ultraviolet-to-radio spectral energy distributions. For comparison we show the redshift distribution for the AS2UDS sample, normalised to match the AS2COSMOS sample size. The AS2COSMOS sample has a median photometric redshift of $z_{\mathrm{phot}}$\,=\,2.68\,$\pm$\,0.06, which is broadly comparable to the median of $z_{\mathrm{phot}}$\,=\,2.61\,$\pm$\,0.04 determined for AS2UDS. However, we note that the AS2COSMOS distribution has a  more extended tail to higher redshifts than is seen in the AS2UDS sample, with 10\,$\pm$\,2\,per cent of AS2COSMOS sources located at $z_{\mathrm{phot}}\geq$\,4 (and 13\,$\pm$\,3 per cent of those brighter than $S_{\rm 870\mu m}=$\,6.2\,mJy) compared to $\sim$\,6\,per cent in AS2UDS. The stacked, normalised probability distribution function (PDF) for all AS2COSMOS sources is shown and is comparable to the median redshift distribution, indicating that our results are not sensitive to asymmetries in the redshift solutions for individual SMGs. {\it Right:} The median photometric redshift of the AS2COSMOS and AS2UDS sources, as a function of their 870\,$\mu$m flux densities. We identify a clear trend of increasing redshift with 870\,$\mu$m flux density, in agreement with results from the AS2UDS survey \citep{Stach18}. We bin the combined AS2COSMOS and AS2UDS surveys by 870\,$\mu$m flux density and show the median of the stacked PDF for all sources in each bin. A linear fit to the median in each bin yields a gradient of 0.06\,$\pm$\,0.01, indicating strong evolution in the average flux density of the SMG population with redshift. 
    }
    \label{fig:photoz}
\end{figure*}

To extract kinematic information from the [C{\sc ii}] line emission from AS2COS\,0001.1\,\&\,0001.2, we first experimented with applying various tapers to the $uv$-data. We found that tapering the data cube to a synthesized beam of 0.4$''$ FWHM provided a good comprise between resolution and surface brightness sensitivity to the line emission. Adopting this tapering strategy, we constructed a ``dirty'' cube for the field that we cleaned following the same procedure used for the AS2COSMOS continuum maps (see \S\,\ref{subsec:datareduc}). The cleaned cube reaches a median sensitivity of 1.0\,mJy\,beam$^{-1}$ per 32\,MHz channel, and we detect line emission from AS2COS\,0001.1\,\&\,0001.2 at an integrated SNR of 32 and 20, respectively (see Figure\,\ref{fig:cii}).

In Figure\,\ref{fig:cii} we show two-dimensional maps of the intensity and kinematics of the  [C{\sc ii}] emission in both SMGs. The kinematic maps were derived from Gaussian fits to the line emission from each source using an adaptive pixel grid; we first consider a single spaxel but, where necessary, adaptively bin pixels to a maximum of 1.5\,$\times$ beam FWHM until we achieve a SNR\,$>$\,5 integrated across the line emission. For both AS2COS\,0001.1\,\&\,0001.2, we see a clear velocity gradient and centrally peaked velocity dispersion that is indicative of bulk, ordered rotation in the line-emitting gas (in contrast \citealt[][]{JimenezAndrade20} only present evidence of ordered
rotation in AS2COS\,0001.1).

The redshift offset we derive between the two ALMA sources corresponds to a  velocity separation of 590\,$\pm$\,40\,km\,s$^{-1}$ (this is marginally higher than the 375\,$\pm$\,50\,km\,s$^{-1}$ derived from the combined 
[C{\sc ii}] and $^{12}$CO(5--4) line kinematics by \citealt[][]{JimenezAndrade20}, but this
difference does not effect the following discussion). 
The two SMGs` have an on-sky separation of 3.1$''$ which, at their estimated redshift, corresponds to a projected spatial separation of $\sim$\,20\,kpc (before accounting for lensing). To understand whether these SMGs are physically-associated we require knowledge of the mass of the dark matter haloes. Clustering measurements of the S2COSMOS sources, and other SMG samples, suggest that typical SMGs at $z$\,$\sim$\,2--3 occupy  dark matter halos of $\sim$\,10\,$^{13}$\,$\Msol$ (\citealt{An19}, see also \citealt{Wilkinson17,Stach20}). Following the discussion in \citet{Wardlow18}, we can expect that pairs of test masses in a  halo with an NFW profile \citep{Navarro97} of mass $\sim$\,10\,$^{13}$\,$\Msol$ have typical velocity seperations of $\sim$\,700\,km\,s$^{-1}$ for a projected spatial separation of $\sim$\,20\,kpc. Hence
the observed spatial and velocity offsets between AS2COS\,0001.1\,\&\,0001.2 are
consistent with them occupying a  dark matter halo which a mass of $\sim$\,10\,$^{13}$\,$\Msol$, suggesting that it is likely these SMGs are physically-associated within a single dark matter halo (see also the discussion 
of S2COSMOS\,0003 in \citealt{Wang16}). Based on the orientation of their
velocity fields, these two galaxies appear to be co-rotating in a prograde orbit, with a velocity offset comparable to their internal rotation velocities, suggesting the possibility we are
witnessing a rapid and
highly efficient merger.  Ths is consistent with the link 
proposed by \cite{Dudzeviciute20} between the SMG population and the highly efficient collapse of gas-rich massive halos, with characteristic masses similar to those inferred here.

\subsection{Redshift distribution and evolution}
\label{subsec:redshifts}

We show in Figure~\ref{fig:photoz} the redshift distribution of the SMGs in our survey derived using the
{\sc magphys} analysis discussed in \S2.6 (see also \citealt{Ikarashi20}).  This shows
the distribution of the median photometric redshift estimated from the PDFs of each source from {\sc magphys}, as well as the summed PDFs, which are in good agreement.  We determine a median redshift for the full sample of AS2COSMOS SMGs  of
$z=$\,2.68\,$\pm$\,0.07, with the subset of sources brighter than our nominal flux limit of $S_{870\mu\rm m}=$\,6.2\,mJy
having a median redshift of $z=$\,2.87\,$\pm$\,0.08.   The latter is marginally higher than the
median of $z=$\,2.61\,$\pm$\,0.08 reported for the somewhat fainter sample of SMGs from AS2UDS by  \cite{Dudzeviciute20}.   Moreover, we note that the AS2COSMOS distribution has a  more extended tail to higher redshifts than is seen in the AS2UDS sample, with 10\,$\pm$\,2\,per cent of AS2COSMOS sources located at $z_{\mathrm{phot}}\geq$\,4 (and 13\,$\pm$\,3 per cent of those brighter than $S_{\rm 870\mu m}=$\,6.2\,mJy) compared to $\sim$\,6\,per cent in AS2UDS.

Similarly, we derive a median redshift of $z=$\,3.24\,$\pm$\,0.19 for the twenty AS2COSMOS SMGs with 870-$\mu$m fluxes
above 12\,mJy, which is marginally higher than the median of $z=$\,2.79\,$\pm$\,0.05 for the 364 SMGs in AS2UDS
brighter than the  3.6\,mJy single-dish flux limit of that study \citep{Dudzeviciute20}.
We can also construct a sample of fourteen SMGs 
with $S_{870\mu\rm m}\leq$\,1\,mJy  from combining AS2COSMOS and AS2UDS,
which have a median redshift of just $z=$\,2.44\,$\pm$\,0.34  (although we caution that around
half of these are secondary components in the maps of brighter SMGs and so may not 
represent an unbiased population, but see the discussion of faint secondary SMGs in \S 3.3).

To better constrain the variation in the redshift of the SMG population with sub-millimetre flux we also show in Figure~\ref{fig:photoz} the correlation between these two parameters for AS2COSMOS combined with the similarly-analysed sample from AS2UDS.  \cite{Stach19} reported a trend between $S_{870\mu\rm m}$ and redshift in the AS2UDS, following earlier suggestions going back over more than two decades \citep{Archibald01,Ivison02,Ivison07}.
We can both test these trends and extend then to higher fluxes using the wider-area and typically brighter AS2COSMOS sample and we see that  Figure~\ref{fig:photoz} indeed shows a strong trend 
of increasing 870-$\mu$m flux with redshift for the combined sample:
we measurement a gradient of the trend in redshift with 870-$\mu$m flux of 0.06\,$\pm$\,0.01\,mJy$^{-1}$ 
for the combined sample, compared to 0.09\,$\pm$\,0.02\,mJy$^{-1}$ estimated from just the AS2UDS sample  in \cite{Stach19}.   This trend between observed 870-$\mu$m flux density and redshift is most likely driven by the increasing gas fraction in these systems and hence gas (and dust) mass in more distant galaxies, compounded by the growing far-infrared
luminosities driven by the higher star-formation rates which are fueled in turn by these more extensive reservoirs of gas 
 \citep{Dudzeviciute20,Ikarashi20}.

\section{Conclusions}

We have presented the results of an ALMA 870-$\mu$m continuum survey of the brightest sub-millimetre  sources drawn from the SCUBA-2 survey of the COSMOS field (S2COSMOS, \citealt[][]{Simpson19,An19}).   Using a combination of our pilot study of 158 
SCUBA-2 sources and comparable observations of a further 24 we construct an effectively complete sample (182/183) of  the sources with
 $S_{850\mu \rm m}\geq $\,6.2\,mJy from the S2COSMOS survey of the 1.6\,deg$^2$ COSMOS field.
The ALMA maps detect 260 SMGs with flux density of $S_{870\mu \rm m}$\,=\,0.7--19.2\,mJy in the 182 fields.  
The main conclusions of this study are:

$\bullet$ We detect multiple SMGs in 34\,$\pm$\,2\ per cent of SCUBA-2 sources, or 53\,$\pm$\,8 per cent for sources brighter than  $S_{850\mu \rm m}>$\,12\,mJy, underlining the fact that blending of
more than one SMG is a significant issue for single-dish surveys.   We estimate that approximately one-third of these SMG--SMG pairs are physically associated, predominantly these are the brighter secondary systems with $S_{870\mu \rm m}\gs$\,3\,mJy). We illustrate these
associated systems using the serendipitous detection of bright [C{\sc ii}] 157.74\,$\mu$m  line emission in the ALMA observations of two  SMGs associated with the highest signal-to-noise SCUBA-2 source
in the field: AS2COS\,0001.1\,\&\,0001.2 at  $z=$\,4.63.

$\bullet$ We show that the number counts derived from our ALMA observations lie below  the raw counts of sources in the S2COSMOS SCUBA-2 survey, but after applying an end-to-end modelling approach which accounts for both source blending
and noise boosting \citep{Simpson19}, the corrected counts from the single-dish survey are in good agreement with those determined from our ALMA observations.  We use this survey and the comparable AS2UDS study of a $\sim$\,1\,deg$^2$ field to derive rough
bounds on the contribution of cosmic variance to the number counts and show these are consistent with predictions from theoretical models.

$\bullet$ We  construct the multiwavelength spectral energy distribution of the AS2COSMOS SMGs using
 the extensive archival data of this field and  model these with {\sc magphys} to estimate their photometric redshifts. We find a median photometric redshift for the $S_{850\mu \rm m}>$\,6.2\,mJy AS2COSMOS sample of $z=$\,2.87\,$\pm$\,0.08, and clear evidence for evolution in the median redshift with 870\,$\mu$m flux density suggesting strong evolution in the bright-end of the 870\,$\mu$m luminosity function.  This is most likely driven by the increasing gas fractions and concomitant high star-formation rates,
and hence dust masses in more distant galaxies \citep{Ikarashi20}.

\section*{Acknowledgements}

The Durham co-authors acknowledge support from STFC (ST/P000541/1) and J.M.S.\ gratefully acknowledges support from the EACOA fellowship programme.
U.D.\ and J.E.B.\ acknowledge the support of STFC studentships (ST/R504725/1 and ST/S50536/1 respectively).  Y.M.\ acknowledges JSPS KAKENHI grant Nos.\ 25287043, 17H04831, and 17KK0098.  W.H.W.\ acknowledges grant support from the
Ministry of Science and Technology of Taiwan (105-2112-M-001-029-MY3 and 108-2112-M-001-014-MY3).  Y.A.\ acknowledges financial support through NSFC grant 11933011. K.E.K.C. acknowledges support from STFC (ST/R000905/1) and a Royal Society/Leverhulme Trust Senior Fellowship (SRF/R1/191013).  This paper makes use of the following ALMA data: ADS/JAO.ALMA\#2016.1.00463.S. ALMA is a partnership of ESO (representing its member states), NSF (USA), and NINS (Japan), together with NRC (Canada), NSC and ASIAA (Taiwan), and KASI (Republic of Korea), in cooperation with the Republic of Chile. The Joint ALMA Observatory is operated by ESO, AUI/NRAO, and NAOJ.
This analysis used data from the S2COSMOS survey (M16AL002) on the JCMT, which in turn 
included data from S2CLS (MJLSC02) and the JCMT archive.
The James Clerk Maxwell Telescope is operated by the East Asian Observatory on behalf of The National Astronomical Observatory of Japan; Academia Sinica Institute of Astronomy and Astrophysics; the Korea Astronomy and Space Science Institute; the Operation, Maintenance and Upgrading Fund for Astronomical Telescopes and Facility Instruments, budgeted from the Ministry of Finance (MOF) of China and administrated by the Chinese Academy of Sciences (CAS), as well as the National Key R\&D Program of China (No.\ 2017YFA0402700). Additional funding support is provided by the Science and Technology Facilities Council of the United Kingdom and participating universities in the United Kingdom and Canada (ST/M007634/1, ST/M003019/1, ST/N005856/1). The James Clerk Maxwell Telescope has historically been operated by the Joint Astronomy Centre on behalf of the Science and Technology Facilities Council of the United Kingdom, the National Research Council of Canada and the Netherlands Organization for Scientific Research and data from observations undertaken during this period of operation is used in this manuscript. This research used the facilities of the Canadian Astronomy Data Centre operated by the National Research Council of Canada with the support of the Canadian Space Agency.


\bibliographystyle{mnras}
\bibliography{ref.bib} 

\appendix
\section{}

\begin{figure*}
	\centering
    \includegraphics[width=1.8\columnwidth]{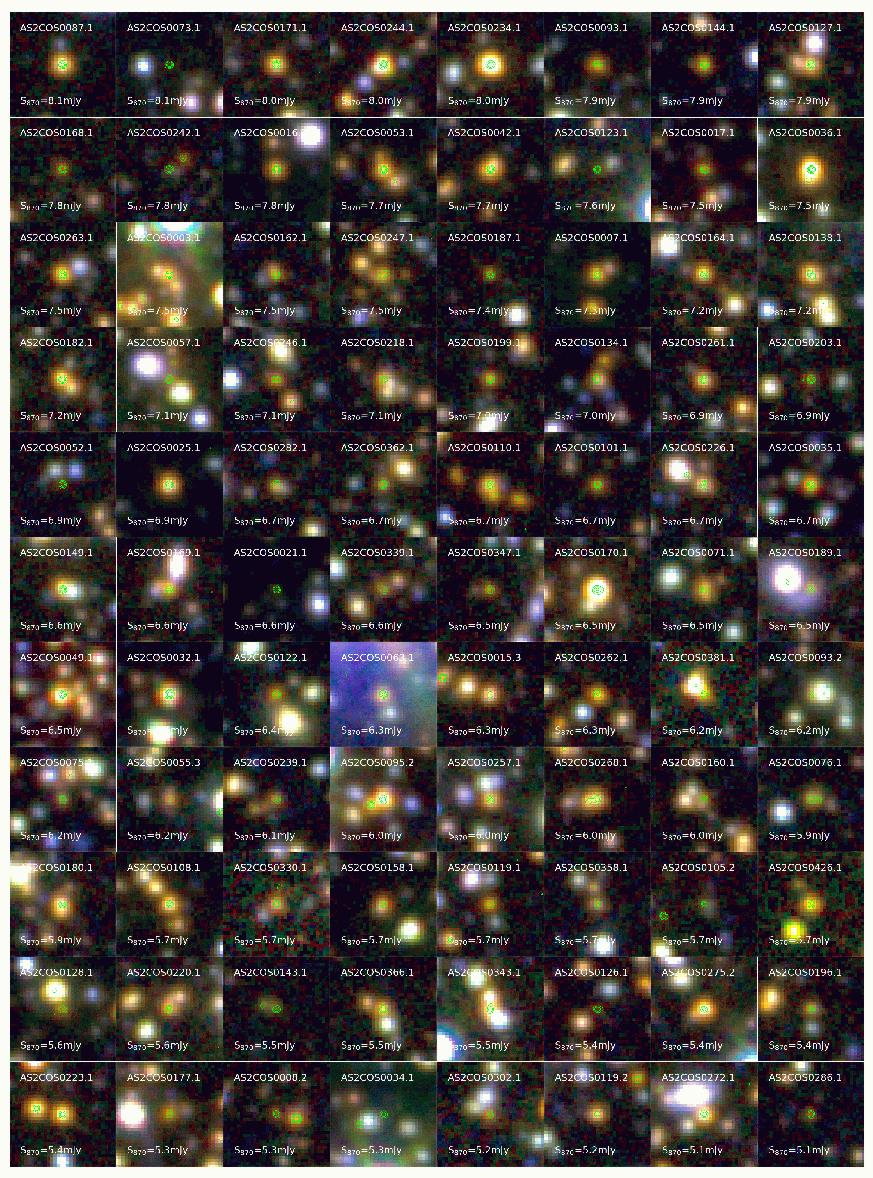}
    \caption{20$''$\,$\times$\,20$''$ true colour images (comprising $K_{s}$, 3.6\,$\mu$m $\&$ 4.5\,$\mu$m) of the 180\,/\,260 ASCOSMOS SMGs with $S_{870\mu\rm m}\leq$\,8.5\,mJy. Each image is centred at the position of the ALMA counterparts, with contours representing 
the 870-$\mu$m emission at 4, 10, 20 and 50\,$\times$\,$\sigma$. Similar colour images for the subset of SMGs that are brighter than 8.5\,mJy are shown in Figure\,\ref{fig:multimages}
    }
    \label{fig:multimages_app}
\end{figure*}

\begin{figure*}
	\centering
    \includegraphics[width=1.8\columnwidth]{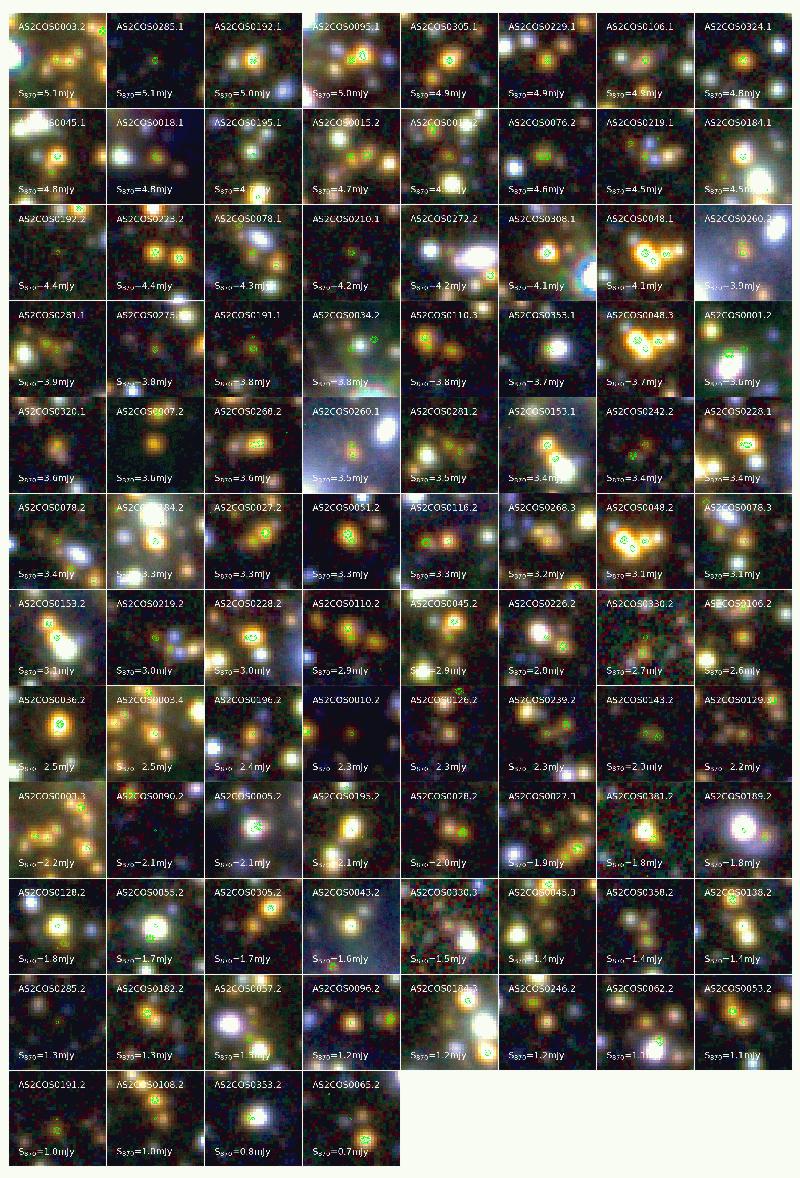}
    \contcaption{}
\end{figure*}


\bsp	
\label{lastpage}
\end{document}